\documentclass[sn-apa]{sn-jnl}

\jyear{2021}%

\theoremstyle{thmstyleone}%
\theoremstyle{thmstyletwo}%

\theoremstyle{thmstylethree}%

\usepackage{supertabular}
\usepackage{siunitx}
\usepackage{soul}
\usepackage{tabu}
\usepackage{makecell}
\usepackage{bm,amsmath}

\newcommand{\pp}[1]{\left(#1\right)}

\newcommand{\dd}[2]{\frac{\partial #1}{\partial #2}}

{}

\begin{document}

\title[An implicit large-eddy simulation perspective on the flow over periodic hills]{An implicit large-eddy simulation perspective on the flow over periodic hills}


\author[1]{\fnm{Laura} \sur{Prieto Saavedra}}\email{laura.prieto-saavedra@polymtl.ca}

\author[1]{\fnm{Catherine E. Niamh} \sur{Radburn}}\email{cenr20@bath.ac.uk}

\author[1]{\fnm{Audrey} \sur{Collard-Daigneault}}\email{audrey.collard-daigneault@polymtl.ca}

\author*[1]{\fnm{Bruno} \sur{Blais}}\email{bruno.blais@polymtl.ca}

\affil*[1]{\orgdiv{Research Center in Industrial Flow Processes (URPEI), Department of Chemical Engineering}, \orgname{Polytechnique Montr\'eal}, \orgaddress{\street{PO Box 6079, Station Centre-Ville}, \city{Montr\'eal}, \postcode{H3C 3A7}, \state{QC}, \country{Canada}}}


\abstract{The periodic hills simulation case is a well-established benchmark for computational fluid dynamics solvers due to its complex features derived from the separation of a turbulent flow from a curved surface. We study the case with the open-source implicit large-eddy simulation (ILES) software Lethe. Lethe solves the incompressible Navier-Stokes equations by applying a stabilised continuous finite element discretisation. The results are validated by comparison to experimental and computational data available in the literature for Re = 5600. We study the effect of the time step, averaging time, and global mesh refinement. The ILES approach shows good accuracy for average velocities and Reynolds stresses using less degrees of freedom than the reference numerical solution. The time step has a greater effect on the accuracy when using coarser meshes, while for fine meshes the results are rapidly time-step independent when using an implicit time-stepping approach. A good prediction of the reattachment point is obtained with several meshes and this value approaches the experimental benchmark value as the mesh is refined. We also run simulations at Reynolds equal to 10600 and 37000 and observe promising results for the ILES approach.
}




\maketitle


\section{Introduction}

The phenomenon of turbulent separation from a curved surface occurs in a large variety of engineering problems, such as flow over the blades of a turbine, past an obstruction in a pipe, and near to an impeller in a mixing tank. It is therefore essential that a method capable of simulating turbulent flows is able to capture this phenomenon and the resulting flow characteristics. The periodic hills is an established simulation benchmark for flow separation \citep{ercoftac2017}. In this case, a well-defined flow passes over a series of hills which repeat along a channel in a periodic fashion. As the flow passes over a hill, there is a pressure-induced separation from the curved surface. It then recirculates on the leeward face of the hill and reattaches at the base of the channel before accelerating up and over the next hill. This case includes complex flow features such as the generation of an unsteady shear layer, recirculation, strong pressure gradients, attached and detached boundary layers, and turbulence recycling due to the periodicity assumption \citep{Gloerfelt2019}.

Over the past decade, the main research focus around the periodic hills simulation case has been on developing better wall functions and subgrid-scale models for explicit Large-Eddy Simulations (LES) (e.g., \cite{Beck2014, Mokhtarpoor2017,Wang2021}), with a few studies using implicit LES (ILES) \citep{Hickel2008,Chen2014,Li2015,Balakumar2015, Krank2018, Wang2021}. In the latter studies, only two use the Finite Element Method (FEM): \cite{Krank2018} with high-order discontinuous FEM and \cite{Wang2021} with hp-spectral-FEM. The main difference between LES and ILES is that in the latter there is no subgrid-scale model. Instead, the refinement of the mesh determines the length scales that are resolved. In general, the mesh is finer in areas of interest or where large flow variation occurs (particularly in near-wall regions), so that the smaller eddies can also be resolved. In stabilised approaches, if the cell size is not small enough to resolve all the eddies up to the scale where there is viscous dissipation, which is usually the case, additional dissipation is included numerically according to what is called a stabilisation term that comprises a stabilisation parameter and the strong residual of the momentum equation.

The aim of this study is two-fold: the first is to demonstrate how accurate results for the periodic hills case can be obtained using less degrees of freedom than a traditional explicit LES approach when using an ILES approach with a stabilised FEM discretisation for different Reynolds numbers. This is achieved by comparing the simulation results to two previous studies: an experimental study by \cite{Rapp2009} and a computational finite volume explicit LES completed by \cite{Breuer2009}. The second aim and main contribution is to investigate the effect of numerical parameters, such as time step, overall simulation time for averaging of flow properties and mesh refinement, on the periodic hills simulation. The solver used in this study is already implemented in the open-source multiphase flow simulation software Lethe, a stabilised continuous Galerkin FEM solver based on the deal.II Library \citep{dealII93, dealII94}. This is the first work in the literature that studies closely the effect of such parameters when using a stabilised FEM approach and it highlights the strengths of stabilised methods for the modeling of these kind of turbulent flows when using an implicit time-stepping scheme. 

The remainder of this work is organized as follows. Section \ref{sec:case} introduces the periodic hills case in detail and summarizes the previous studies available in literature. Section \ref{sec:setup} presents the stabilised formulation, simulation parameters and benchmark data. In Section \ref{sec:results} the results for all the simulations are presented and analysed. Finally, Section \ref{sec:conclusion} summarizes our conclusions.

\section{Periodic hills case} \label{sec:case}
\begin{figure}[ht]
    \centering
    \includegraphics[scale=0.2]{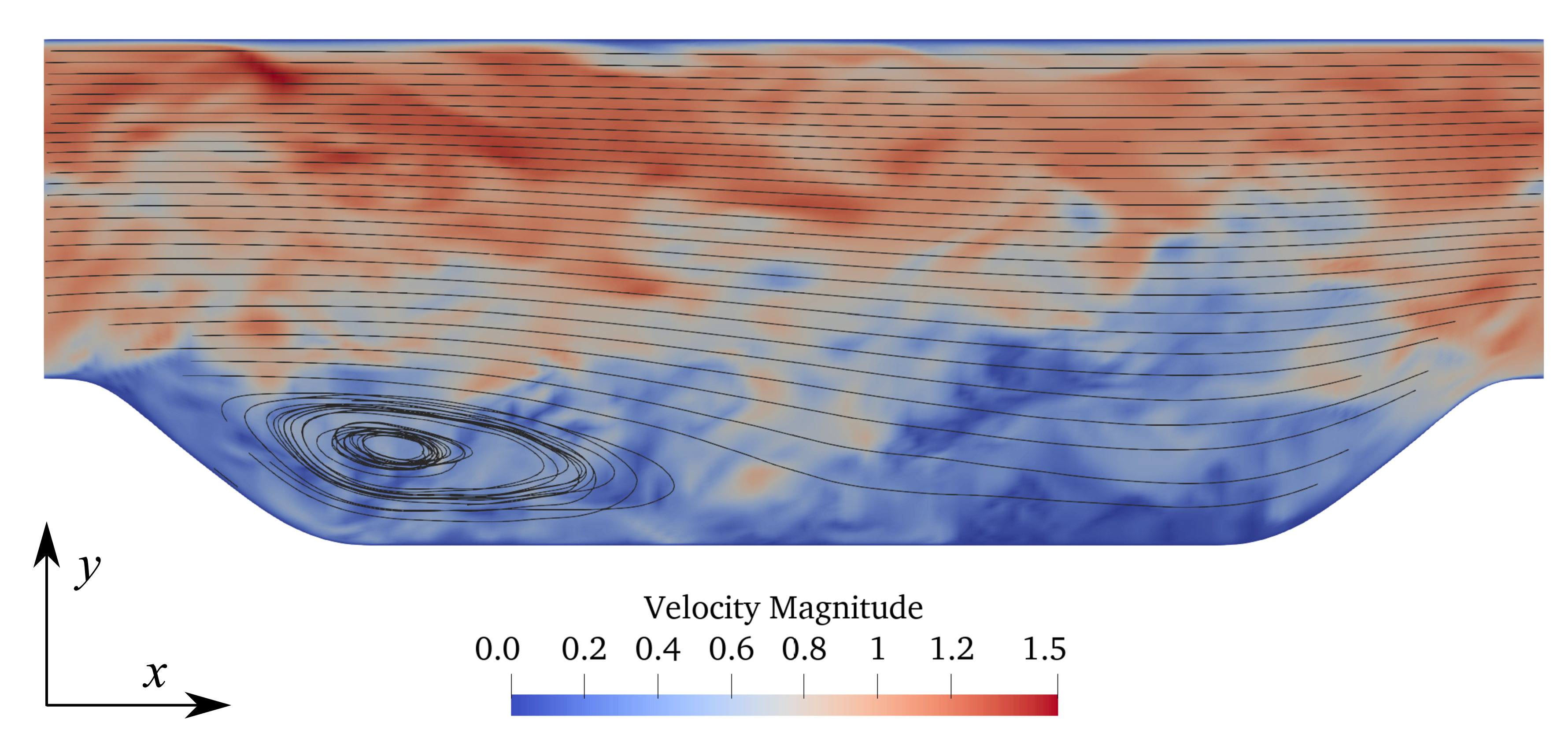}
    \caption{An instantaneous snapshot of the turbulent flow over periodic hills as generated by Lethe after $\SI{800}{\second}$ of simulation run time. The recirculation zone, turbulence shed from the shear layer, and streaks along the upper wall can clearly be seen. The black lines represent streamlines of the time-averaged flow.}
    \label{fig:flow_image}
\end{figure}

As the flow passes over the hill, it is subjected to the effects of both the curvature of the hill and the pressure gradient. The adverse pressure gradient on the leeward side of the hill and resulting deceleration of the flow causes the boundary layer to separate from the curved hill surface. The flow then recirculates on the leeward side of the hill and reattaches in the base of the channel before the next hill. There is a short distance remaining before the subsequent hill which allows the boundary layer to recover. The flow then accelerates up and over the second hill, and the flow pattern repeats in a periodic manner. Fig.~\ref{fig:flow_image} depicts an instantaneous snapshot of the flow over periodic hills.

\subsection{Geometry} \label{subsec:geometry}
\begin{figure}[ht]
    \centering
    \includegraphics[scale=0.6]{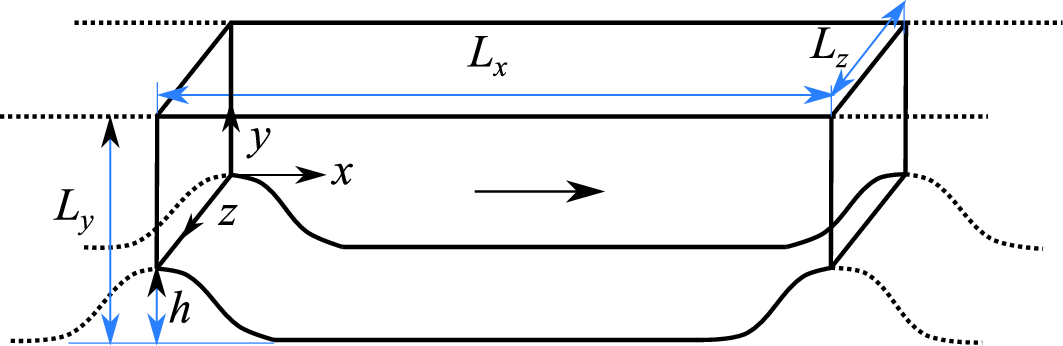}
    \caption{Geometry of the periodic hills test case, where $L_x = 9h$, $L_y = 3.035h$ and $L_z = 4.5h$, and $h$ is the maximum height of the hill (Adapted from \cite{ercoftac2017}).}
    \label{fig:geometry}
\end{figure}

For a couple of decades, many experimental and computational studies have been completed over the same generalised geometry, which was first introduced by \cite{Mellen2000} (for more information on this, a history of the periodic hills case is well surmised by \cite{Manhart2011} and more recently by \cite{Wang2021}). The initial geometry was improved over the years to ensure that the periodic hills case could be used as a benchmark for wall modeling, subgrid-scale modeling and grid parameters \citep{Breuer2009}. For example, the distance between the hills was increased to have a larger reattachment zone and the side walls were eliminated to remove the spanwise effects on the flow. This means that there are now many studies, both experimental and computational, which can be used as benchmarks for the periodic hills case and confirm that the case has a well-defined configuration \citep{Frohlich2005, Rapp2009, Breuer2009, Gloerfelt2015}. Since the majority of cases use the same geometry, it was also logical for our case to use this geometry (see Fig.~\ref{fig:geometry}), allowing comparison of the simulation with both experimental and simulation data.

The shape of the hills are described using 6 polynomials, each defined for a sub-domain of the $x$ domain \citep{ercoftac2017} (the polynomials can be found in Appendix \ref{app:polynomials}). The top of the first hill is located at $x/h = 0$ with an elevation of $y(x) = h$; $y(x)$ reaches a minimal value of 0 at $x/h = 1.929$. The geometry is flat in the range $x \in [1.929; 7.071]$, with the geometry mirrored at $x/h = 4.5$ (meaning the second hill has a windward face equal and opposite to the leeward face of the first hill). The gap between hills is sufficiently sized to allow the flow to reattach between hills and give some distance for recovery of the boundary layer after reattachment. Therefore, the presence of the second hill does not affect the point of reattachment.

The height and width of the channel are specified as to reduce the computational power and memory required and allow sufficient resolution in both directions. The spanwise domain of $L_z = 4.5h$ allows the side wall effects to be ignored and the spanwise fluctuations to be completely resolved excluding the largest eddies \citep{Mellen2000, Frohlich2005, Rapp2009}. Flow characteristics which are half a channel width apart are uncorrelated \citep{Wang2021}.

\subsection{Boundary conditions} \label{subsec:boundary_conditions}
The top and bottom walls use no-slip boundary conditions, while the boundary conditions at the start ($x=0$) and end of the geometry ($x=L_x$) are periodic. This allows the flow regime to achieve periodicity after several flow throughs, as per the definition of the test case, and removes the complication of specifying inlet and outlet conditions. The side walls are considered to have periodic boundary conditions, which allows the model to represent the bulk flow of the channel \citep{ercoftac2017}.

It is worth noting that while the velocity components are explicitly periodic at the streamwise boundaries, the pressure is not. The pressure is comprised of a linear force component and a non-linear pressure component:
\begin{equation} \label{eq:pressure}
\begin{split}
     p_{\mathrm{total}}(\bm{x}, t) = p_{\mathrm{force}}(t) + p_{\mathrm{dynamic}}(\bm{x},t) = \beta(t) x + p(\bm{x},t)
\end{split}
\end{equation}

\noindent where $\beta$ is a spatially independent momentum source term and $p$ is the dynamic pressure resulting from the flow regime. The value of $\beta$ is dynamically adapted through time following the procedure of \cite{Benocci1990} to ensure that the specified volumetric flow rate remains constant as the flow develops.

\subsection{Interesting features} \label{features_of_periodic_hills}
Several features of interest have been identified in the flow pattern at varying Reynolds numbers. The Reynolds number for this flow configuration is given by:
\begin{align} \label{Re}
    \mathrm{Re} &= \frac{u_B h}{\nu}  \\
    u_B &= \frac{1}{2.035 h} \int_{h}^{3.035h} u(y) \mathrm{d}y
\end{align}

\noindent where $u_B$ is the bulk velocity, $\nu$ the kinematic viscosity and $u(y)$ the streamwise velocity profile at $x/h = 0$. Periodic hills simulations have been carried out at Reynolds numbers up to 37000 \citep{Chaouat2013, Mokhtarpoor2016, Mokhtarpoor2017, DelaLlavePlata2018}. Most of the recent LES/DNS periodic hills studies use $\mathrm{Re} = 10595$ (e.g., \cite{Breuer2009, DelaLlavePlata2018, Lodato2019, Krank2018, Wang2021, Gloerfelt2019}). The flow regime is still relevant for lower Reynolds numbers however, with the relevant features being observed at $\mathrm{Re} = 5600$, not to mention the plentiful experimental and numerical data available for this specific case summarized in Section \ref{sec:previous_studies}. Consequently, this Reynolds number is used for the core of this study, while the higher Reynolds numbers are only investigated shortly to have a complete view of the capabilities of the ILES approach. 

One main feature which has a large impact on the overall flow pattern is the point of separation. The point of separation from smoothly contoured walls is affected greatly by the properties of the flow and the environment, including pressure gradient, wall roughness, turbulence in the shear layer and transport of downstream effects \citep{Manhart2011, Balakumar2015}. The point of separation has a strong effect on the length and height of the recirculation bubble, and hence on the reattachment point. As $\mathrm{Re}$ increases, the recirculation zone flattens and the reattachment point moves upstream \citep{Manhart2011}. A small variation in the separation point strongly affects the development of the flow regime in the leeward side of the hill, and so it leads to a major difference in reattachment point \citep{ercoftac2017}. Indeed, it has been shown that a 1\% change in separation point results in a 7\% change in reattachment point \citep{Frohlich2005}.

The accuracy of the reattachment point is a good indicator of the accuracy of the simulation at the near-wall region. However, the turbulent and unsteady nature of the flow leads to a low frequency oscillation of the reattachment point, making it difficult to accurately determine \citep{Manhart2011}. It is usually obtained as the location on the channel flow where $u$ changes direction (the flow stops reversing within the recirculation bubble and reattaches, flowing forward) \citep{Rapp2009}, or the location on the channel floor where the wall shear stress is zero \citep{Hickel2008}.

Large eddies originate from the separated shear layer and are apparent as large longitudinal rolls on the windward side of the second hill \citep{Chaouat2013}. These are due to the Kelvin-Helmholtz effect; the difference in shear stress through the fluid lead to a rotational effect and result in vortex rolls. These elongated vortices are often less than $h$ in diameter, but being able to model these large 3D structures indicates the need to have sufficient spanwise resolution in a simulation.

Fluctuations are observed in the $z$ direction on the windward side of the second hill. This is attributed to the ``splattering" effect. Eddies, including the larger Kelvin-Helmholtz eddies, are transported by convection towards the second hill. These eddies are compressed by the presence of the second hill, and as they can no longer continue motion in the x direction, they ``splatter" outwards. This is much more noticeable in the $z$ direction than in the $y$ direction as the bulk flow at this location has a significant $y$-component as it accelerates up and over the second hill \citep{Frohlich2005}. 


A small recirculation has also been identified at the foot of the second hill, at around $x/h = 7$. As the flow travels over the second hill, the flow accelerates and shear stress increases rapidly.  Flow at the base of the second hill decelerates and can reverse, leading to a separation of the boundary layer and secondary vortices \citep{Frohlich2005}. However, this separation is strongly dependent on the flow regime before the second hill and transportation of eddies. Since this varies over time, this recirculation is hardly visible in the averaged flow field \citep{Breuer2009}.

Another potential small recirculation can be observed on the crest of the hill. The sudden increase of wall shear stress as the flow accelerates over the second hill is countered by a change from a favourable to an adverse pressure gradient at approximately $x/h = 8.6$, dropping the shear stress and decelerating the flow. This deceleration can lead to a small flat recirculation zone (only found in numerical simulations at $\mathrm{Re}>10600$, e.g., \cite{Breuer2009, diosady2014}). Hence, it is not expected to be observed in the Lethe simulations at $\mathrm{Re} = 5600$. For a more detailed review of the flow features which arise in the periodic hills simulation case, we refer the reader to dedicated previous studies such as the articles by \cite{Frohlich2005} or by \cite{Gloerfelt2015}.

\subsection{Previous studies of the periodic hills simulation case} \label{sec:previous_studies}


Table \ref{tab:meshes} describes the type and resolution of meshes used in previous periodic hills simulations at $\mathrm{Re} = 5600$, and the resulting reattachment points obtained, to give an indicator of the near-wall resolution. Similarly, Table \ref{tab:previous_studies_2} and Table \ref{tab:previous_studies_3} contain the information for studies using $\mathrm{Re} = 10600$ and $\mathrm{Re} = 37000$, respectively. Table \ref{tab:meshes} shows that the mesh size varies dramatically across studies of periodic hills at $\mathrm{Re} = 5600$. Meshes used for LES tend to vary from 4 million cells up to 218 million cells. The spacing and number of grid points is important to resolve features within the flow completely and accurately. The finer the mesh, the closer to DNS the simulation becomes, and the greater the computational expense. Several studies do compare mesh size and then proceed with the most accurate mesh for the least computational expense \citep{Chaouat2013, Krank2018}. 

\begin{sidewaystable*}
    \sidewaystablefn%
    \begin{center}
    \caption{Mesh type and resolutions used in previous studies with $\mathrm{Re} = 5600$.}
    \label{tab:meshes}
    \begin{minipage}{\textheight}
    \resizebox{\textheight}{!}{%
    \begin{tabular}{@{\extracolsep{\fill}}c c c c c c c c c c@{\extracolsep{\fill}}} 
    \toprule
    \textbf{Code} & \textbf{Type} & \textbf{Spatial method}  & \makecell{\textbf{Number of} \\ \textbf{cells}} & \textbf{DoFs} & \textbf{Mesh type} & \makecell{\textbf{Reattachment} \\ \textbf{point}} & \makecell{\textbf{Averaging time}\\\textbf{[flows through]}} & \makecell{\textbf{Time step}\\\textbf{[s]}} & \textbf{Author}             \\ 
    \midrule
    LESOCC        & LES           & \makecell{FVM \\ (cell-centered)}                     &    $4.7$M   &   -      & Curvilinear        & 4.56          & 55  & 0.002   & \makecell{\cite{Frohlich2005}}       \\ [1ex] 
    STREAMLES     & LES           & \makecell{FVM \\ (cell-centered)}                     &    $4.7$M   &   -      & Curvilinear        & 4.72          & 55  & 0.001   & \makecell{\cite{Frohlich2005}} \\ [1ex] 
    LESOCC Case 7 & LES           & FVM                     & $12.4$M &  -       & Curvilinear        & 5.09          & 145 & 0.002   & \makecell{\cite{Breuer2009}}         \\ [1ex] 
    MGLET Case 8  & DNS           & FVM                     & $218$M &  -        & Cartesian          & 5.14          & 38  & 0.001   & \makecell{\cite{Breuer2009}}         \\ [1ex] 
    URDNS 5600    & URDNS         & \makecell{DG-FEM \\ $(k = 6)$}      & $65$K & $22.5$M & Curvilinear        & 4.82$\pm$0.09 & 61  & Dynamic & \makecell{\cite{Krank2018}}          \\ [1ex] 
    DNS 5600      & DNS           & \makecell{FEM \\ $(k = 7)$}        & $65$K & $33.6$M & Curvilinear        & 5.04$\pm$0.09 & 61  & Dynamic & \makecell{\cite{Krank2018}}          \\ [1ex] 
    Incompact3d   & DNS           & FD                     &   $37.8$M    &   -      & Cartesian          & $\sim$4.70    & 150 & 0.0005  & \makecell{\cite{Xiao2020}}   \\
    \bottomrule
    \end{tabular}}
    \footnotesize \footnotetext{\textbf{Note:} A dash (-) is placed where the information was not explicitly reported in the respective reference.}
    \end{minipage}
    \end{center}
\end{sidewaystable*}

\begin{sidewaystable*}
    \sidewaystablefn%
    \begin{center}
    \caption{Mesh type and resolutions used in previous studies with $\mathrm{Re} = 10600$.}
    \label{tab:previous_studies_2}
    \begin{minipage}{\textheight}
    \resizebox{\textheight}{!}{%
    \begin{tabular}{@{\extracolsep{\fill}}c c c c c c c c c c@{\extracolsep{\fill}}} 
    \toprule
    \textbf{Name} & \textbf{Type} & \textbf{Spatial method} & \makecell{\textbf{Number of} \\ \textbf{cells}} & \textbf{DoFs} & \textbf{Mesh type} & \makecell{\textbf{Reattachment} \\ \textbf{point}} & \makecell{\textbf{Averaging time}\\\textbf{[flows through]}} & \makecell{\textbf{Time step}\\\textbf{[s]}} & \textbf{Author}             \\ 
    \midrule
    ALDM            & ILES  & \makecell{FVM \\ (staggered)} &    4.5M  & -  & Cartesian        & 4.3 & 60   &  - & \makecell{\cite{Hickel2008}}          \\ [1ex]
    LESOCC Case 9  & LES           & FVM                  & $12.4$M & - & Curvilinear        & 4.69          & 142 & 0.0018   & \makecell{\cite{Breuer2009}}\\ [1ex]
    WMLES\_C   & ILES           & FVM      & $200$K & - & Cartesian & -          & 90 & Dynamic   & \makecell{\cite{Chen2014}}\\ [1ex]
    WMLES\_F   & ILES           & FVM      & $900$K & - & Cartesian & -          & 40 & Dynamic   & \makecell{\cite{Chen2014}}\\ [1ex]
    MDCD/SLAU   & ILES           & FVM                  & $900$K to $6.9$M & - &     Curvilinear    & - & 20 & $0.01$   & \makecell{\cite{Li2015}}\\
    High order impact   & \makecell{LES \\ ILES}           & FVM  & $5.4$M and $14.3$M & - &     Curvilinear    & \makecell{4.2 and 4.4 \\ 3.75 and 4.4}          & -  & -   & \makecell{\cite{Balakumar2015}}\\ [1ex]
    DRP11   & LES           & FD                  & $4.2$M $33.5$M & - &     Curvilinear    & $-$ & 80 & $-$   & \makecell{\cite{Gloerfelt2015}}\\  [1ex]       
     No-model and WALE      & LES  & \makecell{DG-FEM \\ (k=3)}  &  $65$K  & $4.19$M  & Curvilinear        & 3.9 & -  & 0.0001 & \makecell{\cite{DelaLlavePlata2018}}          \\ [1ex]
    URDNS 10600    & URDNS         & \makecell{DG-FEM \\ (k=5)}       &  $524$K  & $113$M  & Curvilinear        & 4.57$\pm$0.06 & 61  & Dynamic & \makecell{\cite{Krank2018}}          \\ [1ex]
    DNS 10600      & DNS           & \makecell{FEM \\ (k=6)}          &   $524$K  & $180$M  & Curvilinear        & 4.51$\pm$0.06 & 61  & Dynamic & \makecell{\cite{Krank2018}}          \\ [1ex]
    FD    & LES  &   FDM    & \makecell{$67$K, $524$K, \\ $4.2$M and $33.5$M} & -  & Curvilinear        & - &  55 to 80 & Explicit & \makecell{\cite{Gloerfelt2019}}        \\ [1ex]
    SEDM-Roe    & LES  &   \makecell{DFEM \\ (k=4)}    &  $72$K and $10$K & $9$M and $1.3$M  & Curvilinear        & 4.37 and 4.18 &  64 and 96 & Explicit & \makecell{\cite{Lodato2019}} \\ [1ex]
    SEDM-Aufs    & LES  &   \makecell{DFEM \\ (k=4)}    &  $72$K and $10$K & $9$M and $1.3$M  & Curvilinear        & 4.21 and 4.43 &  64 and 96 & Explicit & \makecell{\cite{Lodato2019}} \\ [1ex]
    SEDM    & LES  &   \makecell{DFEM \\ (k=6)}   &  $72$K & $24.7$M  & Curvilinear        & - &  23 & Explicit & \makecell{\cite{Lodato2019}}   \\ [1ex]
     Nektar$++$   & ILES   & \makecell{hp-FEM \\ (k=4,7)}    &  $246$K, $500$K  & -  & Curvilinear        & - & 140  & 0.001 & \makecell{\cite{Wang2021}}          \\
    \bottomrule
    \end{tabular}}
    \footnotesize \footnotetext{\textbf{Note:} A dash (-) is placed where the information was not explicitly reported in the respective reference.}
    \end{minipage}
    \end{center}
\end{sidewaystable*}

\begin{sidewaystable}
    \sidewaystablefn%
    \begin{center}
    \caption{Mesh type and resolutions used in previous studies with $\mathrm{Re} = 37000$.}
    \label{tab:previous_studies_3}
    \begin{minipage}{\textheight}
    \resizebox{\textheight}{!}{%
    \begin{tabular}{@{\extracolsep{\fill}}c c c c c c c c c c@{\extracolsep{\fill}}} 
    \toprule
    \textbf{Name} & \textbf{Type} & \makecell{\textbf{Spatial}\\\textbf{method}} & \makecell{\textbf{Number of} \\ \textbf{cells}} & \textbf{DoFs} & \textbf{Mesh type} & \makecell{\textbf{Reattachment} \\ \textbf{point}} & \makecell{\textbf{Averaging time}\\\textbf{[flows through]}} & \makecell{\textbf{Time step}\\\textbf{[s]}} & \textbf{Author} \\
    \midrule
    \makecell{HHTBLEC \\ HHTBLEF}   & ILES           & FVM  & $200$K and $350$K & $12.4$M &     -    & 3.41 and 3.80          & 90 and 40 & -   & \makecell{\cite{Chen2010}}\\ [1ex]
    PITM       & \makecell{Hybrid RANS\\ /LES}  & FVM & \makecell{$240$K, $480$K, \\ $480$K and $960$K}    &    -     & Curvilinear        & \makecell{4.30, 4.26 \\ 3.54 and 3.68}          & -  & Explicit       & \makecell{\cite{Chaouat2013}} \\[1ex]
    LES & LES  & FVM   &  \makecell{500K, 5M \\ and 20M}  & -  & Curvilinear        &  \makecell{3.5, 3.65 \\ and 3.65} & 140  & Explicit & \makecell{\cite{Mokhtarpoor2016}} \\ [1ex]
    DLUM & RANS/LES  & FVM  &   500K  &  - & Curvilinear        &  3.8 & 140  & Explicit & \makecell{\cite{Mokhtarpoor2016}} \\ [1ex]
    WALE       & LES  & \makecell{DG-FEM \\ (k=3)} & $65$K   &  $4.19$M & Curvilinear        & 3.2 &  - & Explicit & \makecell{\cite{DelaLlavePlata2018}}\\
    \bottomrule
    \end{tabular}}
    \footnotesize \footnotetext{\textbf{Note:} A dash (-) is placed where the information was not explicitly reported in the respective reference.}
    \end{minipage}
    \end{center}
\end{sidewaystable}

The majority of previous periodic hills simulations do not discuss or even state the time averaging or time stepping aspects of the simulations, excluding whether to consider if the CFL number is sufficiently low (an important consideration for stability in explicit time stepping methods). If they do state the averaging time and time step, often these are extremely large and small respectively. However, an overly large averaging time or overly small time step may result in a simulation running for much longer and being much more computationally costly than necessary. Additionally, due to the transient and turbulent nature of the simulation, it was theorised that these two parameters would have an effect on the results produced. Therefore, the time after which the average was taken and the time step used in the simulation were deemed to be two parameters worthy of further investigation.

To the authors' knowledge, the only two FEM ILES studies that study some of these parameters for the periodic hills case are by \cite{Krank2018} and \cite{Wang2021}. The time is considered by the former, who look at the time taken for averages to converge for different DNS and under resolved DNS simulations with both $\mathrm{Re} = 5600$ and $\mathrm{Re} = 10600$. They also consider the convergence using \textit{h/p} refinement for a discontinuous Galerkin ILES approach, while the latter look at the ILES requirements only for $\mathrm{Re} = 10600$, but mainly focus on the grid requirements and order of the elements, and do not consider the effect of time at all.


In summary, the majority of previous studies of the periodic hills simulation case have focused on the physical processes controlling the flow regime, or solely have demonstrated the ability of a CFD code to accurately model the simulation case, rather than focused on the parameters of the simulation itself. They hence ran the simulation case at the smallest time step and finest grid that is computationally feasible for the longest time.

Optimisation of the parameters for the flow regime has barely been focused on at all; using parameters which maintain accuracy of the simulation while reducing computational expense are important for feasible simulations, particularly for application of the model to industry \citep{Kornhaas2008}. Therefore, the effect of the numerical parameters must be understood and so the effect of time step, averaging time and mesh resolution, are investigated within this work.

\section{Simulation setup} \label{sec:setup}
All simulations in this work were completed using the open-source software Lethe \citep{blais2020lethe}. All relevant code required to run the simulations in this report can be found in the public Github repository for Lethe (\url{https://github.com/lethe-cfd/lethe.git}). Code specific to post-processing the results obtained from the periodic hills simulations is available in the periodic hills folder within the Lethe-utils Github repository (\url{https://github.com/lethe-cfd/lethe-utils.git}). Lethe is dependent on the deal.II library v9.4.0 with Trilinos, p4est and MPI enabled \citep{dealII93, dealII94}. In this study,  the supercomputers Beluga and Niagara of the Digital Research Alliance of Canada were used.

\subsection{Governing equations and numerical model}

Lethe solves the incompressible Navier-Stokes equations:
\begin{align}
\nabla \cdot \bm{u} &= 0 \label{eq::ns_1} \\
\dd{\bm{u}}{t}+\pp{\bm{u}\cdot \nabla} \bm{u} &=  -\nabla p^* + \nabla \cdot \bm{\tau} +\bm{f}
\label{eq::ns_2}
\end{align}
with 
\begin{align}
\bm{\tau} = \nu \pp{\pp{\nabla \bm{u}} + \pp{\nabla \bm{u}}^T}
\end{align}

\noindent where $\bm{u}$ is the velocity vector, $p^*=\frac{p}{\rho}$ with $p$ the pressure and $\rho$ the density, $\bm{\tau}$ the deviatoric stress tensor, $\nu$ the kinematic viscosity and $\bm{f}$ a body force.  Being non-linear partial differential equations, they must be discretised in space and time in order to approximate a solution. For this a continuous Galerkin Finite-Element formulation is used along with a SUPG (Streamline-Upwind/Petrov-Galerkin)/PSPG (Pressure-Stabilizing/ Petrov-Galerkin) stabilization approach. This allows the use of equal order finite elements for the pressure and the velocity components and avoids numerical oscillations for advection-dominated problems \citep{Tezduyar1992,Ilinca2019,blais2020lethe}. The following weak formulation for the Navier-Stokes equations is obtained:
\begin{align}
&\int_{\Omega} \nabla \cdot \bm{u} q d\Omega 
+ {\sum_K \int_{\Omega_k} \pp{\dd{\bm{u}}{t}+\bm{u}\cdot\nabla\bm{u}+\nabla p^* - \nabla \cdot \bm{\tau} - \bm{f}} \cdot 
\pp{\tau_{u} 
\nabla q} d\Omega_k}=0 \label{eq::NS_gls1}
\\
&\int_{\Omega}  \pp{\dd{\bm{u}}{t}+\bm{u}\cdot \nabla \bm{u} - \bm{f}} \cdot \bm{v} d\Omega + \int_{\Omega} \bm{\tau} : 
\nabla \bm{v} 
d\Omega - \int_{\Omega} p^* \nabla \cdot \bm{v} d\Omega \nonumber \\&+ {\sum_K \int_{\Omega_k} 
\pp{\dd{\bm{u}}{t}+\bm{u}\cdot\nabla\bm{u}+\nabla p^* - \nabla \cdot \bm{\tau} - \bm{f}} \cdot \pp{\tau_{u}\bm{u} \cdot \nabla 
\bm{v}} 
d\Omega_k} = 0 \label{eq::NS_gls2}
\end{align}

\noindent where $\bm{v}$ and $q$ are the test functions for velocity and pressure respectively, and $K$ is the total number of elements. Since the problem is transient, the stabilization parameter $\tau_{u}$ takes the following form:
\begin{align}
\label{tau_transient}
\tau_u = \left[ \pp{\frac{1}{\Delta t}}^2 + \pp{\frac{2 \|\bm{u}\|}{h_{\mathrm{conv}}}}^2 + 9 \pp{\frac{4\nu}{h_{\mathrm{diff}}^2}}^2 \right]^{-1/2}
\end{align}

\noindent where $\Delta t$ is the time step, $h_{\mathrm{conv}}$ and $h_{\mathrm{diff}}$ are the size of the element related to the convective transport and diffusion mechanism, respectively \citep{ Tezduyar1992d, Tezduyar1992}. In Lethe, both element sizes ($h_{\mathrm{conv}}$ and $h_{\mathrm{diff}}$) are set to the diameter of a sphere of a volume equivalent to that of the cell \citep{ Blais2018,Ilinca2019, blais2020lethe}. The full definition of the methods and capability of this CFD solver are detailed in a separate publication  \citep{blais2020lethe}. For the simulations in this study, Newton's method is  used to solve implicitly the non-linear problem. Each linear system of equations is solved with an ILU preconditioned GMRES solver. A second-order backward difference implicit scheme (BDF2) is used for time-stepping \citep{Hay2015}.


\subsection{Simulation parameters and mesh}

The flow geometry is used as described in Section \ref{subsec:geometry} and the boundary conditions as specified in Section \ref{subsec:boundary_conditions}. The height of the hill ($h$) is set equal to $1$ for simplicity. Likewise, the volumetric flow rate and kinematic viscosity are set to be $\SI{9.1575}{\meter \cubed \per \second}$ and $\SI{1.78571E-04}{\meter \squared \per \second}$ respectively, so that the bulk velocity is $u_B =$ $\SI{1}{\meter \per \second}$, and a Reynolds number of $5600$ can be maintained.  The kinematic viscosity is set to $\SI{9.43396E-05}{\meter \squared \per \second}$ for a Reynolds number of $10600$ and to $\SI{2.7027E-05}{\meter \squared \per \second}$ for a Reynolds number of $37000$. For all the simulations, Lagrange elements of first order ($Q_1$) are used for both pressure and velocity. 

\begin{figure}[ht]
    \centering
    \includegraphics[scale=0.2]{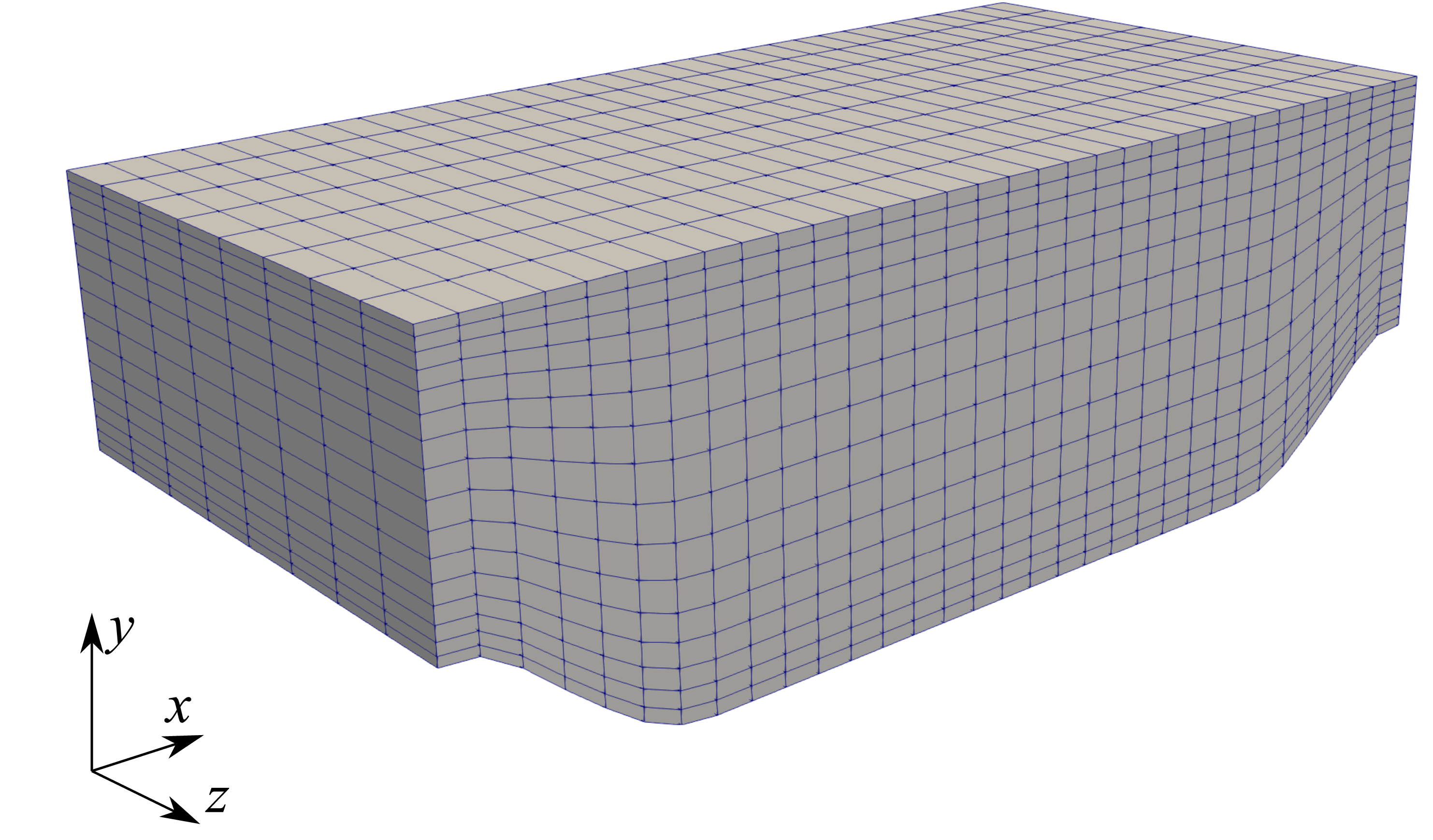}
    \caption{Example of the curvilinear mesh used in Lethe. This mesh contains only 4K cells for the purpose of visualization. However, the coarsest mesh used in the results section contains 120K cells.}
    \label{fig:mesh}
\end{figure}

The elements of the mesh are isoparametric hexahedra and are arranged on a curvilinear mesh (see Fig~\ref{fig:mesh}). The mesh used for all the simulations is a static, uniformly refined mesh. In the $x$ and $z$ directions, the elements are of equal width across the domain. In the $y$ direction, spacing of the grid points can be varied, allowing the mesh to become finer as the elements approach the wall. For all the meshes used in this study, the ratio between the longest to the shortest dimensions of a cell located in the middle of the geometry never exceeds a value of two.

The quality of the grid is as important as the number of grid points in order to accurately locate features of interest and reduce numerical errors. The resolution of the mesh in the near-wall region is evaluated using the dimensionless distance from the wall $y^+$ which gives an indication of how fine the mesh is in the near-wall region:
\begin{equation} \label{y+}
\begin{split}
    y^+ = \frac{y_{cc} u_\tau}{\nu}
\end{split}
\end{equation}

\noindent where $y_{cc}$ is the half of distance from the wall to the wall-nearest grid point, $\nu$ is the kinematic viscosity and $u_\tau$ is the friction velocity given by:
\begin{equation} \label{u_tau}
\begin{split}
    u_\tau = \sqrt{\frac{\tau_w}{\rho}}
\end{split}
\end{equation}

\noindent where $\tau_w$ corresponds to the wall-shear stress and $\rho$ to the density.  The spanwise and lengthwise cell lengths $\Delta x^+$ and $\Delta z^+$ are also important for resolving near the wall. However, if $\Delta y^+$ is satisfied for the simulation in question, the cell shape implies that the requirements for $\Delta x^+$ and $\Delta z^+$ are almost always satisfied too \citep{Frohlich2005, Gloerfelt2019}. Apart from the $y^{+}$ criteria, that is in fact commonly used in the LES and RANS domains, no other criteria in the literature of stabilised methods were found to assess the quality of the mesh a priori. 

The $y^+$ distribution for all the meshes used in this study are reported in Appendix~\ref{app:y_plus}. According to the literature, simulations of attached boundary layers are not precise for traditional LES approaches if the nearest computed values are not located within the viscous sub-layer ($y^+<5$) \citep{John2016}. However, no value is to be found for ILES approaches. Despite this, all the meshes for the simulations at $\mathrm{Re} = 5600$ and $\mathrm{Re} = 10600$ have an average value lower than $y^+=3.4$. In the case of the meshes used for $ \mathrm{Re} = 37000$, the coarsest mesh has the highest average value equal to $7.06$.

\subsection{Comparison data} \label{comparison_data}

In order to verify the accuracy, stability and reliability of this ILES approach in the periodic hills case, the results are compared to established test data from both experiments and other CFD simulations. Experimental data is obtained from \cite{Rapp2009}, and results from the LESOCC CFD code performed by \cite{Breuer2009} provide benchmark computational data. For the latter, the data was extracted from the article using Engauge digitizer \citep{mitchell2017engauge}. Since data is provided for cross-section at varying $x/h$ values in both benchmarks, the results from the Lethe simulation must also be extracted at these points to allow comparison of the data.

\subsubsection{Experimental data}
A series of experiments ran by \cite{Rapp2009} provide an experimental benchmark case. Not all benchmark data can be obtained from a physical experiment - notably the data obtained by Rapp does not contain values of the turbulent kinetic energy or the spanwise Reynolds normal stresses. The experimental set up involved a series of 10 hills with the curvature and hill spacing described in Section \ref{subsec:geometry} with hill height $h = \SI{50}{\milli\metre}$. The channel height was kept the same, but the channel span was increased to $18h$ in order to sufficiently neglect side wall effects.

Results were collected using Particle Image Velocimetry (PIV) and verified against Laser Doppler Anemometry (LDA) measurements. Piezoelectric pressure probes were used to allow non-intrusive pressure measurement. Results were corrected for error and signal noise. These conscientious, well-considered measurement techniques used give high confidence to the results. The reattachment point was determined at $x/h = 4.83$.

Periodicity of the flow was confirmed by comparing flow between hills 6 and 7 and hills 7 and 8, and the reference data taken from between hills 7 and 8.  While it is confirmed that the flow can be assumed to be homogeneous (statistically 2D), the periodicity at $\mathrm{Re} = 5600$ can only be proven to a certain extent due to limitations inherent to the measurement equipment. However, \cite{Rapp2009} concluded that the data produced is precise enough to be used to develop better LES models.

Rapp also collected data for higher Reynolds numbers ($\mathrm{Re} = 10600$ and $\mathrm{Re} = 37000$) using the same set up and experimental techniques. This data is chosen for comparison with our results obtained using the higher Reynolds numbers in Section \ref{subsec:highreynolds}. Measuring properties near to the wall is specially challenging in these cases as the boundary layer thickness is very small, however, the periodicity of the flow could be proven completely in both cases \citep{Rapp2009}. The experimental reattachment point was determined to be equal to $x/h = 4.21$ and to $x/h = 3.76$ for $\mathrm{Re} = 10600$ and $\mathrm{Re} = 37000$, respectively.

\subsubsection{Computational data}
An established benchmark test case for computational periodic hills simulations was created by \cite{Breuer2009} by use of the FVM code LESOCC. LESOCC solves the incompressible Navier-Stokes equations with the sub-grid scale Smagorinsky-Lilly model with Van Driest damping near the solid walls, followed by averaging and filtering to stabilise the dynamic model. A range of simulations were performed by Breuer et al.; the first results of interest arise from Case 7, which was performed at $\mathrm{Re} = 5600$ using a LES approach. This setup uses around 12.4 million active cells (13.1 million grid points) in the mesh with a time step of $\SI{0.002}{\second}$ and the average taken over a time period of $\SI{1300}{\second}$.

The reattachment point for the benchmark case is $5.09$, which is longer than the value of $4.83$ given experimentally by Rapp. Since the reattachment point describes the overall performance of a simulation in one number \cite{Krank2018}, these values can be used as indicators for the method's performance. Breuer et al. note that the pressure distribution in the computational benchmark data is slightly under-predicted due to the SGS model, leading to a delay in the separation point, and hence to an over-prediction of the reattachment point. In general, previous implicit LES studies appear to give better reattachment point agreement to the Rapp data than the Breuer data (e.g., \cite{Wang2021}), but in the bulk of the flow the profiles are closer to those of Breuer data (e.g., \cite{Krank2018}).  Therefore, while both benchmarks do not give precise values, they give a clear indication of the region the reattachment point should fall within.

The second results of interest are known as LESOCC Case 9 and correspond to the Reynolds number of $\mathrm{Re}=10600$. The grid used for this case was the same as in the previously explained simulation, but a time step of $\SI{0.0018}{\second}$ and an average taken over a time period of $\SI{1300}{\second}$ were used. The numerical reattachment value was determined to be $x/h = 4.69$ in this case.

\section{Results and Discussion} \label{sec:results}
A baseline simulation is initially run to validate the results against the computational and experimental data sets presented in Section \ref{comparison_data}. Then, different meshes are considered when investigating the effects of time step and averaging time, followed by some simulations using higher Reynolds numbers. In Table~\ref{tab:simulation_summary}, a summary of the different simulations along with their parameters is presented.


\begin{table}[h]
\begin{center}
\begin{minipage}{\textwidth}
\caption{Summary of the simulations performed and respective parameters.}
\label{tab:simulation_summary}
 \begin{tabular*}{\textwidth}{@{\extracolsep{\fill}}lllll@{\extracolsep{\fill}}}
 \toprule
 \textbf{Case} & \textbf{Mesh} & \textbf{Re} & \textbf{Time step} $[\SI{}{\second}]$ & \textbf{Time average} $[\SI{}{\second}]$ \\ 
\midrule
\multirow[t]{2}{2cm}{Baseline} & Coarse & 5600 & 0.1 & 1000\\ 
& Regular  &  &  &  \\ 
& Fine  &  &  &  \\ [1ex]
\multirow[t]{2}{2cm}{Time step} & Coarse & 5600 & 0.0125, 0.025, & 1000 \\ 
& Regular & & 0.05, 0.1 & \\ 
& Fine & & & \\ [1ex]
\multirow[t]{2}{2cm}{Time averaging} & Coarse & 5600 & 0.025 & 500 to \\ 
&  Regular &  &  & 1100 \\
&  Fine &  &  &  \\ [1ex]
\multirow[t]{2}{2cm}{High Reynolds} & Very coarse & 10600 & 0.025 & 800 \\ 
 numbers &  Coarse & 37000  &  &  \\ 
  &  Intermediate & &  & \\ [1ex]
\bottomrule
\end{tabular*}
\footnotetext{\textbf{Note:} (cells, DoFs) of each mesh: very coarse ($\sim$120K, 500K), coarse ($\sim$250K, 1.1M), intermediate ($\sim$500K, 2.4M), regular ($\sim$1M, 4.5M) and fine ($\sim$4M, 16M).}
\footnotetext[1]{$k$ corresponds to the order of the element.}
\end{minipage}
\end{center}
\end{table}

\subsection{Baseline} 
\label{baseline}

The baseline simulation ensures that the results accurately reproduce the physical phenomenon occurring in the periodic hills case. It uses the coarse, the regular and the fine mesh, a time step of $\SI{0.1}{\second}$, and the average is taken between $\SI{207}{\second}$ and $\SI{1000}{\second}$ (for an averaging period of $\SI{793}{\second}$ or 88 flow throughs). This averaging period is larger than most of the simulations shown in Table~\ref{tab:meshes} and is studied in detail in  Section~\ref{subsec:timeaveraging}.

\begin{figure}[ht]
\centering
\includegraphics[scale=0.7]{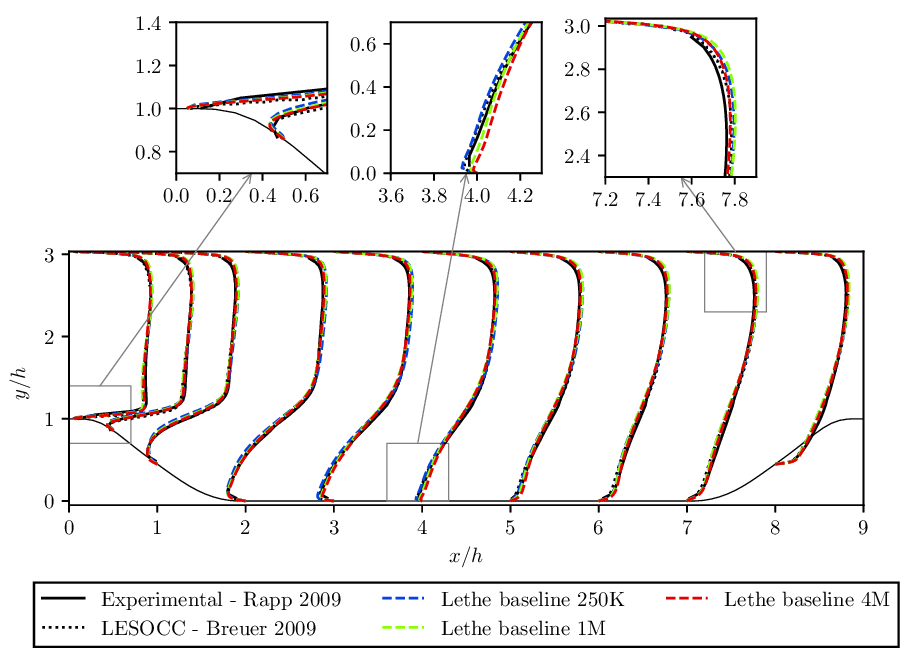}
\caption{Average velocity in the $x$ direction throughout the geometry at $\mathrm{Re} = 5600$, compared against the benchmarks. The profiles are scaled by a factor of 0.8 for ease of visualisation.}
\label{fig:baseline_validation_velocity}
\end{figure}

\begin{figure}[ht]
\centering
\includegraphics[scale=0.7]{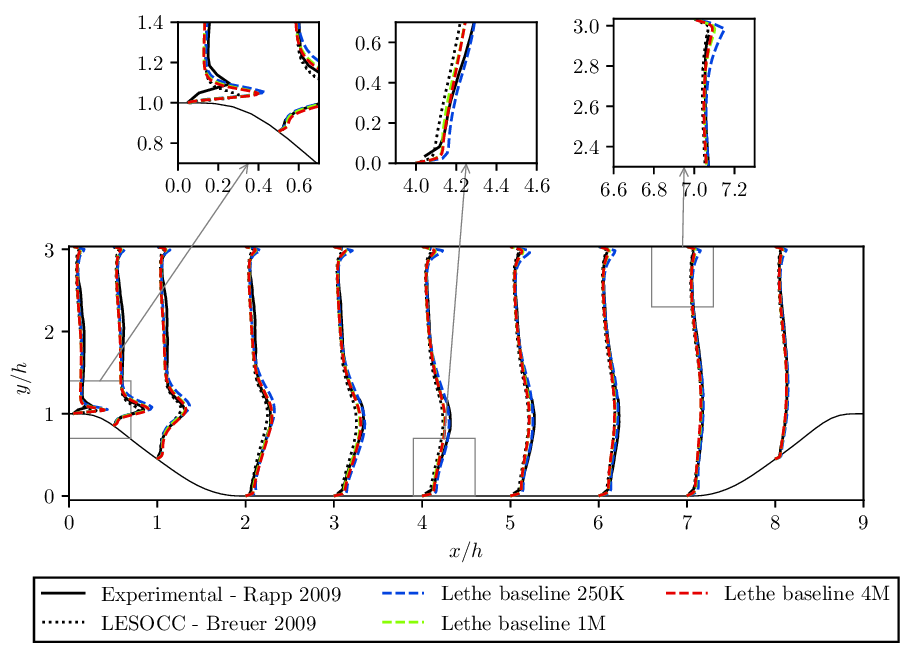}
\caption{Reynolds normal stress in the $x$ direction throughout the geometry at $\mathrm{Re}= 5600$, compared against the benchmarks. The profiles are scaled by a factor of 5 for ease of visualisation.}
\label{fig:baseline_validation_reynolds}
\end{figure}

Considering the average velocity profile in the $x$ direction (Fig.~\ref{fig:baseline_validation_velocity}), there is a good agreement of the Lethe data with both benchmarks in the bulk of flow. At the upper and lower walls, the Lethe data exceeds the benchmarks but retains shape at all $x$ values similarly to the LESOCC simulation. The $y$ velocity and Reynolds stresses also agree well with the benchmark data; for the Reynolds normal stress in $x$ direction see Fig.~\ref{fig:baseline_validation_reynolds}. The Reynolds stresses are more sensitive than the average velocity but overall, there is a good accuracy of the prediction with minimum discrepancies between the meshes. 

The reattachment point was determined to be 4.73 with the coarse mesh, 4.40 with the regular mesh, and 4.35 with the fine mesh; all of them shorter than both the Rapp and the Breuer values (4.83 and 5.09 respectively). In the literature, the reattachment point has been shown to be constantly under-predicted by overly coarse meshes \citep{Wang2021}, however, as we are using a stabilised method, in this case other parameters are playing an important role as well, such as the time step and the averaging period for the estimated quantities. Therefore, these parameters along with the mesh are investigated in the following sections.

\subsection{Time step} \label{subsec:timestep}

To study the effect of the time step in the simulation, four simulations per mesh were completed using Lethe. Again, the coarse, the regular and the fine meshes were used, and the time steps were defined by sequentially halving the value of the time step from $\Delta t = \SI{0.1}{\second}$ to $\Delta t = \SI{0.0125}{\second}$. Average results were taken again after $\SI{1000}{\second}$.

\begin{figure}[ht]
\centering
\includegraphics[scale=0.42]{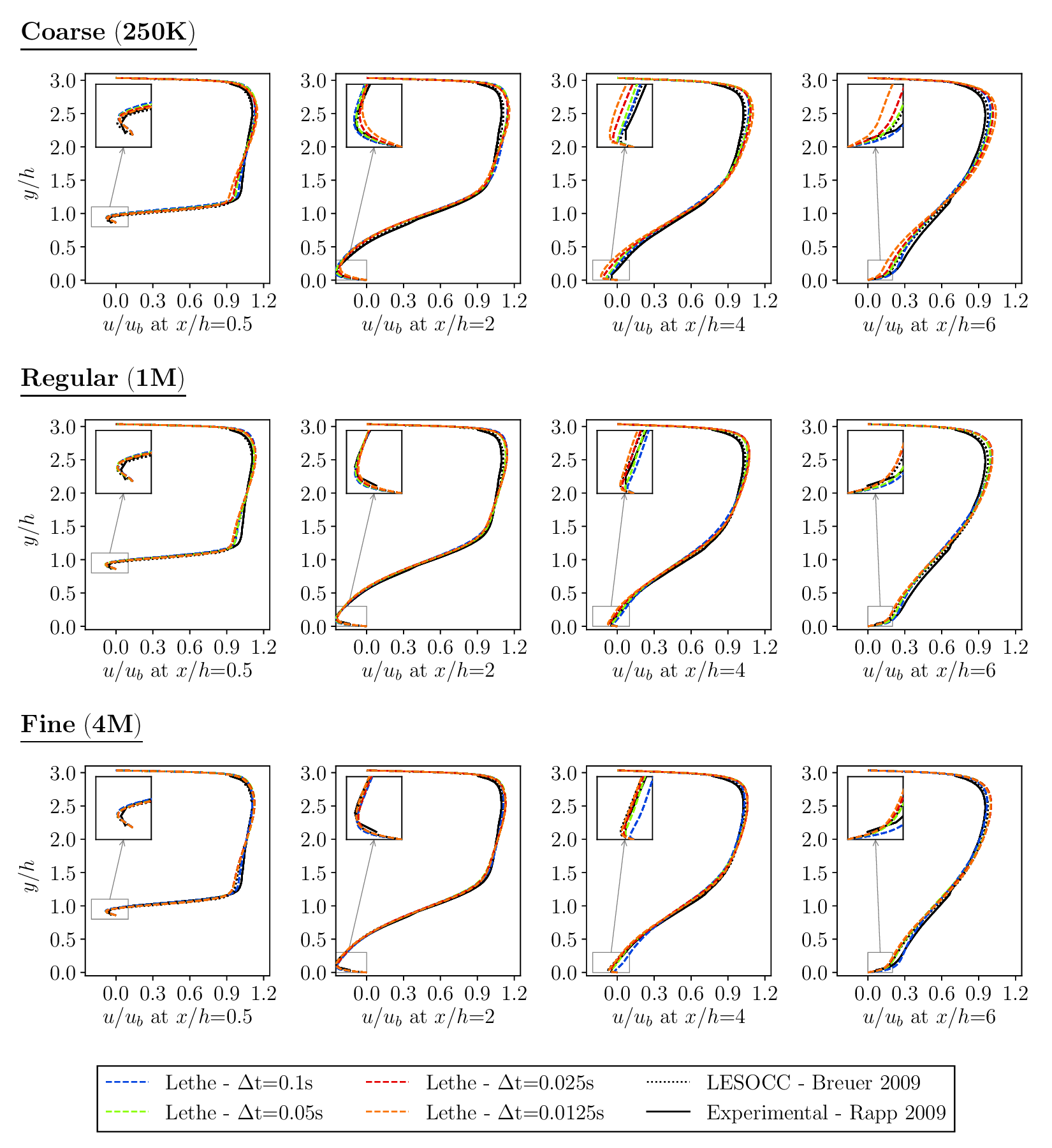}
\caption{Average velocity in the $x$ direction at different points of the geometry with $\mathrm{Re}= 5600$, compared against the benchmarks.}
\label{fig:time_stepping_velocity_0}
\end{figure}

\begin{figure}[ht]
\centering
\includegraphics[scale=0.42]{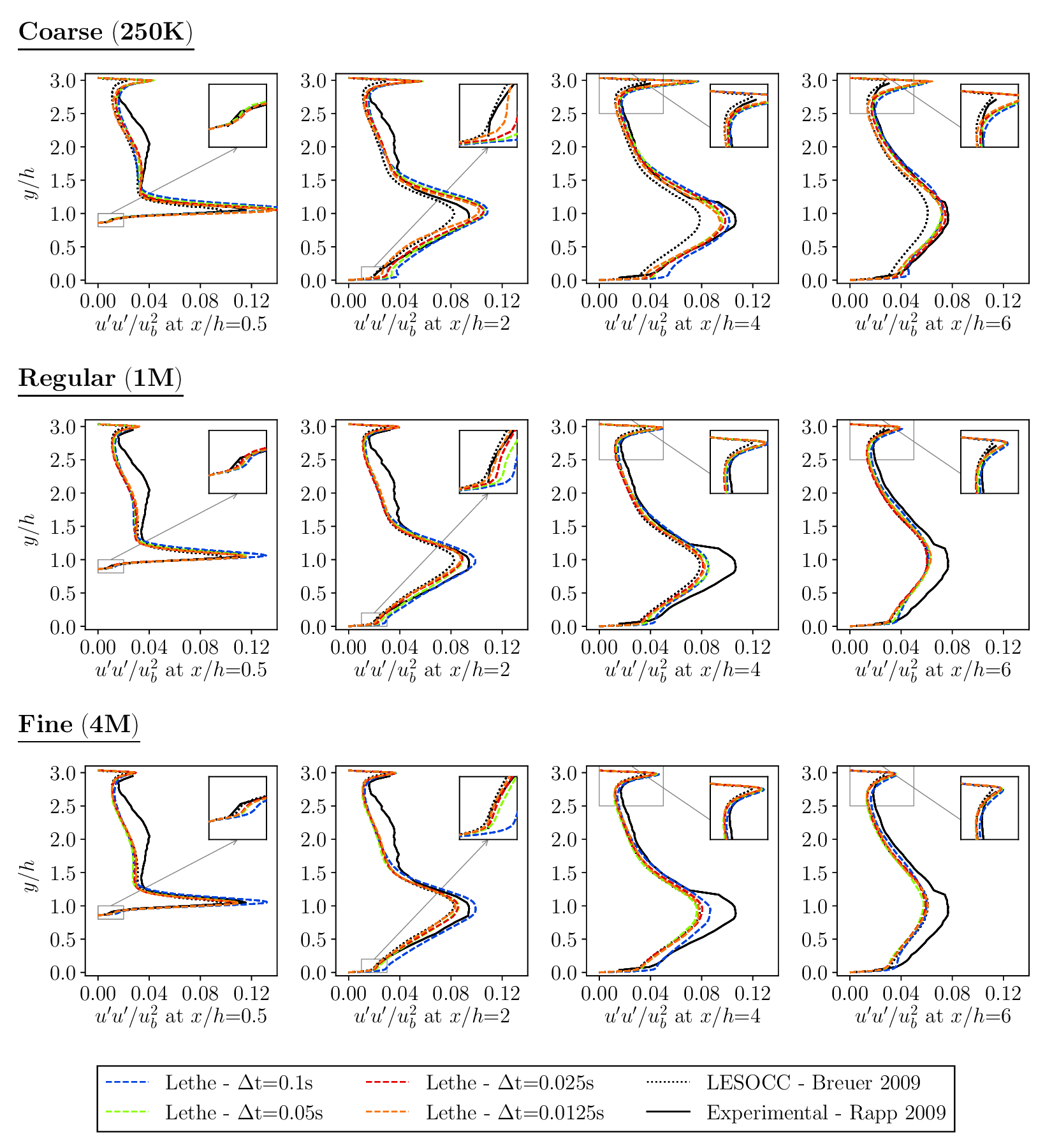}
\caption{Reynolds normal stress in the $x$ direction at different points of the geometry with $\mathrm{Re}= 5600$, compared against the benchmarks.}
\label{fig:time_stepping_reynolds_uu}
\end{figure}

\begin{figure}[ht]
\centering
\includegraphics[scale=0.42]{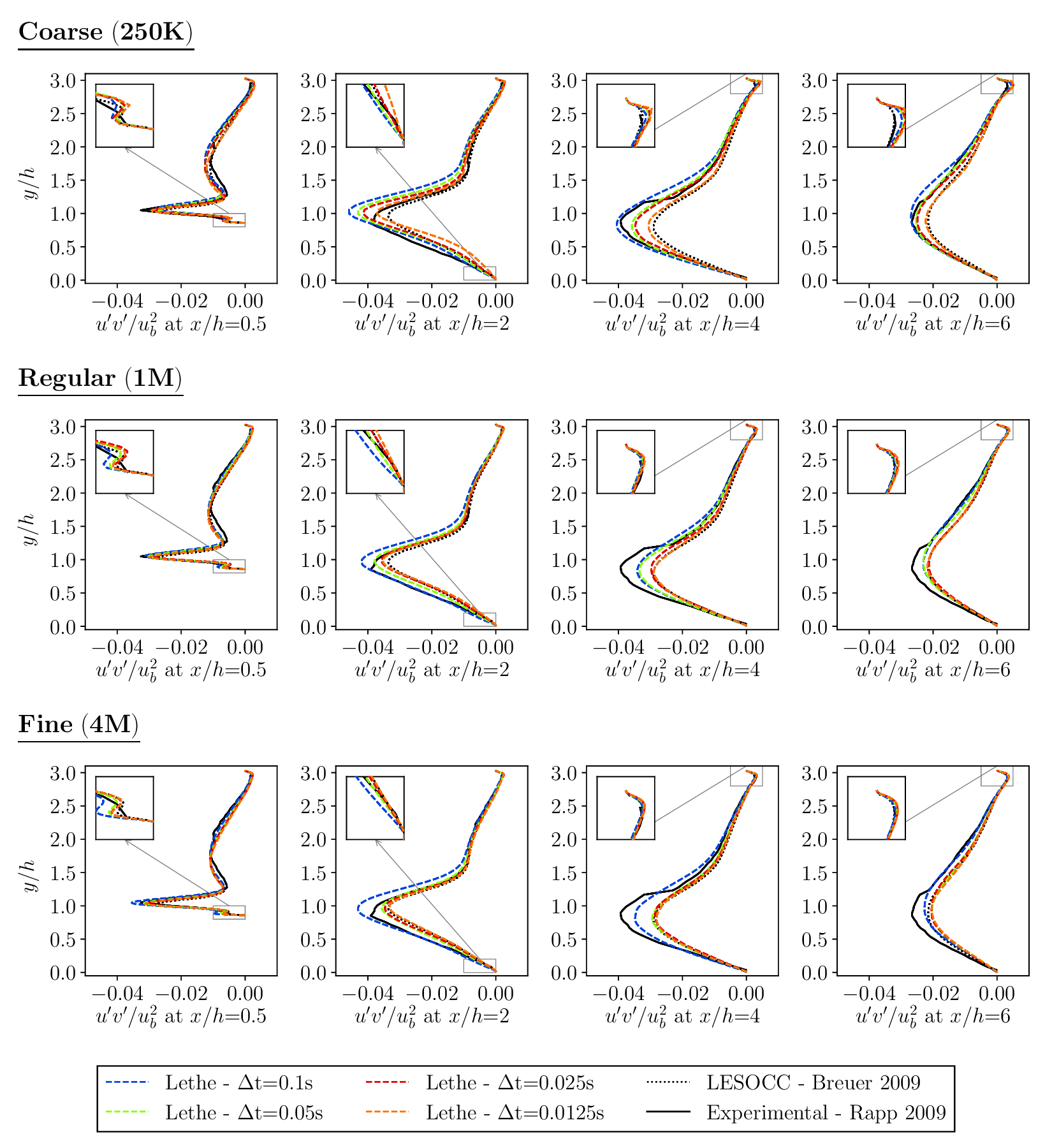}
\caption{Reynolds shear stress at different points of the geometry with $\mathrm{Re}= 5600$, compared against the benchmarks.}
\label{fig:time_stepping_reynolds_uv}
\end{figure}

The average velocity profiles in $x$ are presented in Fig.~\ref{fig:time_stepping_velocity_0}. For all the meshes, we obtain results that are very similar to the experimental and numerical benchmarks. The biggest difference can be observed again in the near-wall region. Looking at the zoom-in plots, it is possible to observe that in the case of the coarse mesh, there is a high discrepancy between the results corresponding to the different time steps. While for the regular and fine mesh, this difference between the results is reduced. The stresses demonstrate this trend most significantly (see Fig.~\ref{fig:time_stepping_reynolds_uu} and \ref{fig:time_stepping_reynolds_uv}). In the case of the fine mesh, all the time steps with exception of the largest one converge towards the numerical solution of \cite{Breuer2009}.

These results can be analyzed by taking into account the stabilisation term in the FEM formulation, which comprises two components: i) the stabilisation parameter $\tau$, which in turn considers the time step $\Delta t$ and the cell size $h$, and ii) the residual of the strong form of the momentum equation. In the case of the coarse mesh, as the time step is reduced, so is the dissipation or stabilisation, which in general leads to a deterioration of the accuracy in coarse meshes. This phenomenon has been observed in the literature, e.g., in the articles by \cite{Hsu2010}, \cite{Calderer2013} and \cite{Gamnitzer2010}. In the case of the fine mesh, the norm of the strong residual is smaller, which reduces the effect of the stabilisation, and leads to accurate and similar results for all the time steps. For the case where $\mathrm{Re} = 5600$, we observe that using a stabilised formulation along with an implicit scheme, allows us to refine the mesh and use a time step as large as $0.05$ without losing the accuracy of the solution. 

\begin{table}[h]
\begin{center}
\begin{minipage}{\textwidth}
\caption{Near-wall parameters and CFL at varying time steps for an averaging time of $\SI{1000}{\second}$. The experimental reattachment points are: 4.83 for the experimental benchmark by \cite{Rapp2009} and 5.09 for the numerical benchmark by \cite{Breuer2009}.}
\label{tab:timestep_parameters}
\begin{tabular*}{\textwidth}{@{\extracolsep{\fill}}lcccccc@{\extracolsep{\fill}}}
\toprule%
& \multicolumn{2}{@{}c@{}}{\textbf{Coarse (250K)}} & \multicolumn{2}{@{}c@{}}{\textbf{Regular (1M)}} & \multicolumn{2}{@{}c@{}}{\textbf{Fine (4M)}} \\\cmidrule{2-3}\cmidrule{4-5} \cmidrule{6-7}%
\textbf{Time step} & \textbf{RP} & \textbf{CFL} & \textbf{RP} & \textbf{CFL} & \textbf{RP} & \textbf{CFL} \\
\midrule
0.1 & 4.73±0.05 & $\approx$ 2.2 & 4.40±0.05 &  $\approx$ 3.2 & 4.35±0.4 & $\approx$ 5.1\\
0.05 & 4.87±0.05 & $\approx$ 0.9 & 4.59±0.05 &  $\approx$ 1.6 & 4.74±0.4 & $\approx$ 2.8 \\
0.025 & 5.07±0.05 & $\approx$ 0.5 & 4.80±0.05 & $\approx$ 0.7 & 4.83±0.4 & $\approx$ 1.2\\
0.0125 & 5.37±0.06 & $\approx$ 0.2  & 4.88±0.05 & $\approx$ 0.4 & 4.85±0.4 & $\approx$ 0.6\\  
\botrule
\end{tabular*}
\footnotetext{\textbf{Note:} RP corresponds to reattachment point.}
\end{minipage}
\end{center}
\end{table}

Considering the reattachment points in Table \ref{tab:timestep_parameters}, it can be seen that the value of the reattachment point increases as time step decreases when using the coarse mesh. For the regular mesh, decreasing the time step means that the value converges more towards the experimental value (4.83) by \cite{Rapp2009}, as the stabilization is affected by the refinement of the mesh. Hence the reattachment point becomes more accurate as the time step decreases. Finally, in the case of the fine mesh, all the time steps, apart from the coarsest one, obtain a nearer value to the one obtained experimentally.

In this study, we use an implicit time-stepping scheme where the CFL condition is not necessary, allowing a much greater time step to be used stability-wise compared to an explicit time stepping scheme. For stability with an explicit time stepping method, a CFL number less than 0.8 is required in the periodic hills case as reported by \cite{Mokhtarpoor2017}, which explains why the time step used is so small in other studies, which mostly use explicit methods. We report the CFL number of each simulation in Table \ref{tab:timestep_parameters}. It is worth noting that for the fine mesh, only the smallest time step fulfills the usual CFL requirement for this case, while all the others do not but still produce accurate results. This indicates that it is possible to use fine meshes and large time steps, without affecting the accuracy, if a stabilized approach with implicit time-stepping scheme is used. A detailed comparison between explicit and implicit time-stepping schemes, in terms of computational cost of the solution is out of the scope of this study. However, it is important to keep in mind that iterations are generally computationally more demanding when using an implicit scheme.

In conclusion, for coarse meshes, reducing the time step leads to a reduction of the accuracy of the average velocity and Reynolds stresses, with a higher impact in the latter for the bulk of the flow. In the case of fine meshes, a similar accuracy is obtained for all the time steps with exception of the coarsest one, which leads to the possibility of using high CFL values when simulating complex cases. 

\subsection{Time averaging} \label{subsec:timeaveraging}

For the periodic hills simulation, the average velocities and Reynolds stresses are taken over time. The values used in the averaging process are taken after the time exceeds $\SI{207}{\second}$, or after 23 hills have been passed, and so the flow can be considered as periodic. For the reference numerical data, the averaging period used was of 145 flow-through times while in the experimental set up a total of 10 hills were considered. In previous studies, the averaging time period varied by approximately one order of magnitude \citep{Krank2018}, with no clear consistency on how that time period was decided. In fact, \cite{Krank2018} stated that the averaging period suggested by some studies for this case is of around $1000$ flow-through times, which is very long and not feasible. Therefore, we decided to study the effect of the time elapsed on the convergence of the average values. For this purpose, we used again the coarse, the regular and the fine mesh and data was extracted at varying times to see the convergence. The lowest averaging period considered was of $\SI{500}{\second}$ or $44$ flow throughs and the largest averaging period was $\SI{1000}{\second}$ or $88$ flow throughs. To be able to have a fair comparison of the effect of the averaging time for these three meshes, a time step of $\Delta t = \SI{0.025}{\second}$ is chosen due to the observations of the previous section.

\begin{figure}[ht]
\centering
\includegraphics[scale=0.42]{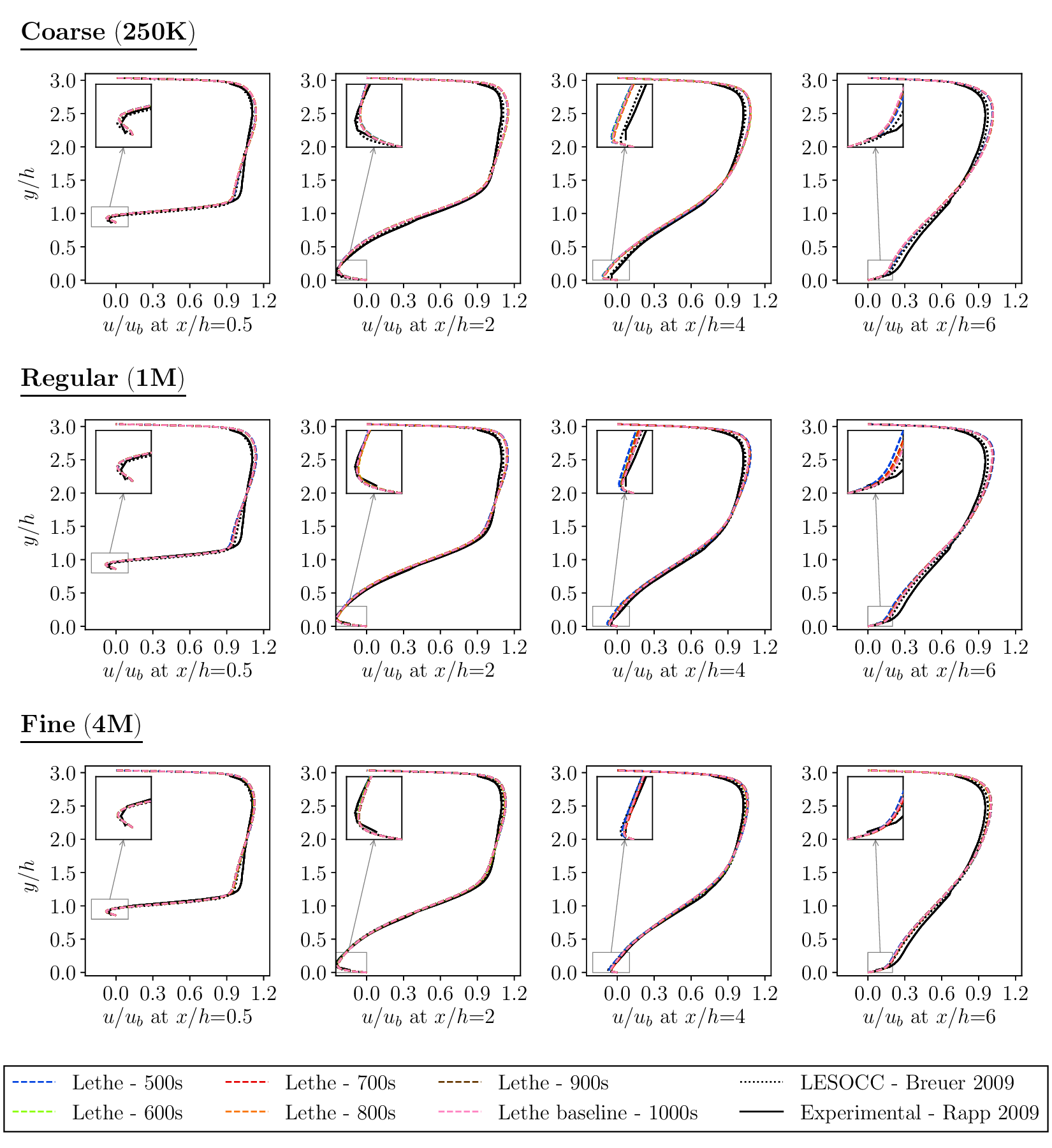}
\caption{Average velocity in the $x$ direction at different points of the geometry with $\mathrm{Re}= 5600$, compared against the benchmarks.}
\label{fig:time_averaging_velocity_0}
\end{figure}

\begin{figure}[ht]
\centering
\includegraphics[scale=0.42]{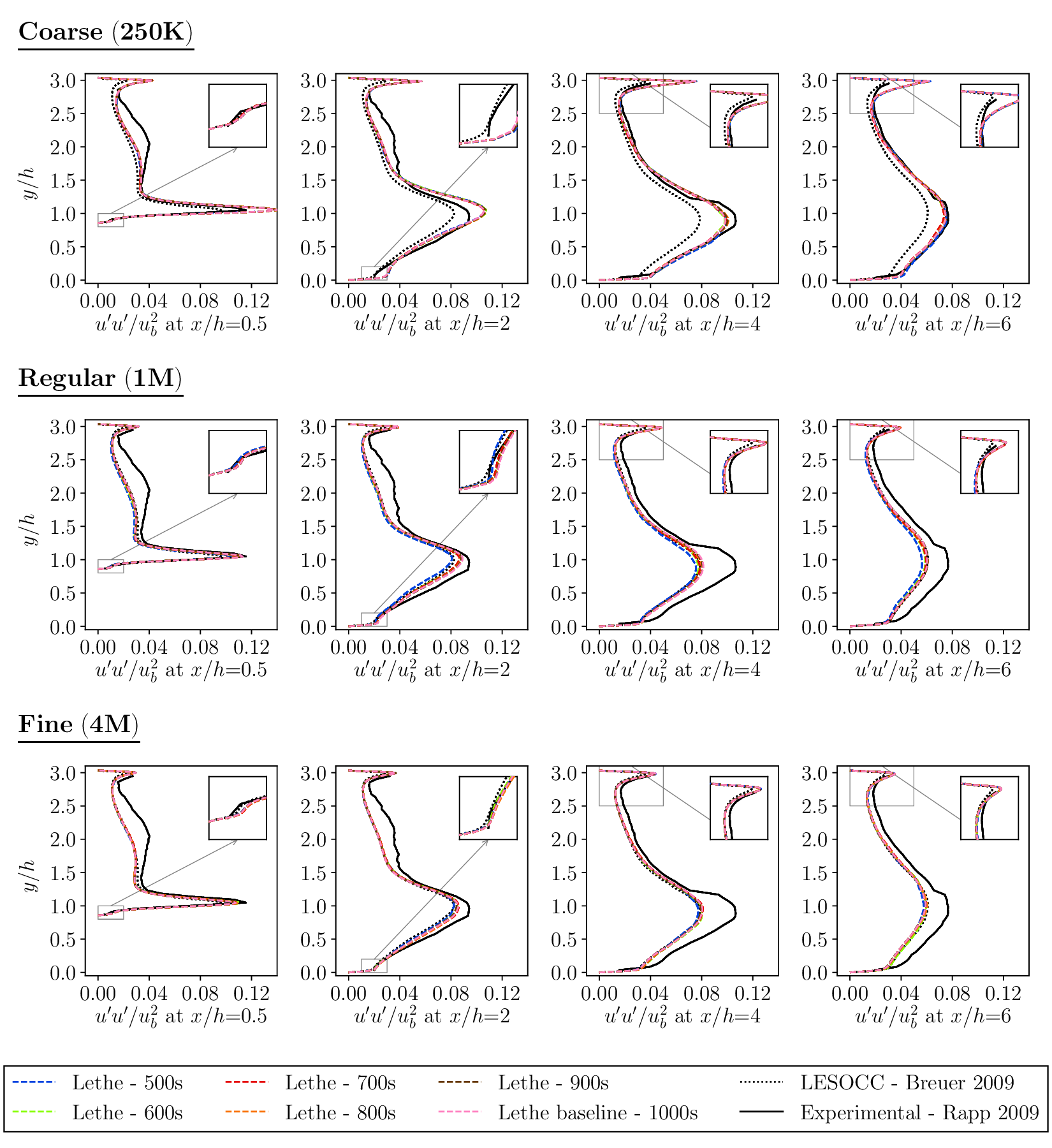}
\caption{Reynolds normal stress in the $x$ direction at different points of the geometry with $\mathrm{Re}= 5600$, compared against the benchmarks.}
\label{fig:time_averaging_reynolds_uu}
\end{figure}

\begin{figure}[ht]
\centering
\includegraphics[scale=0.42]{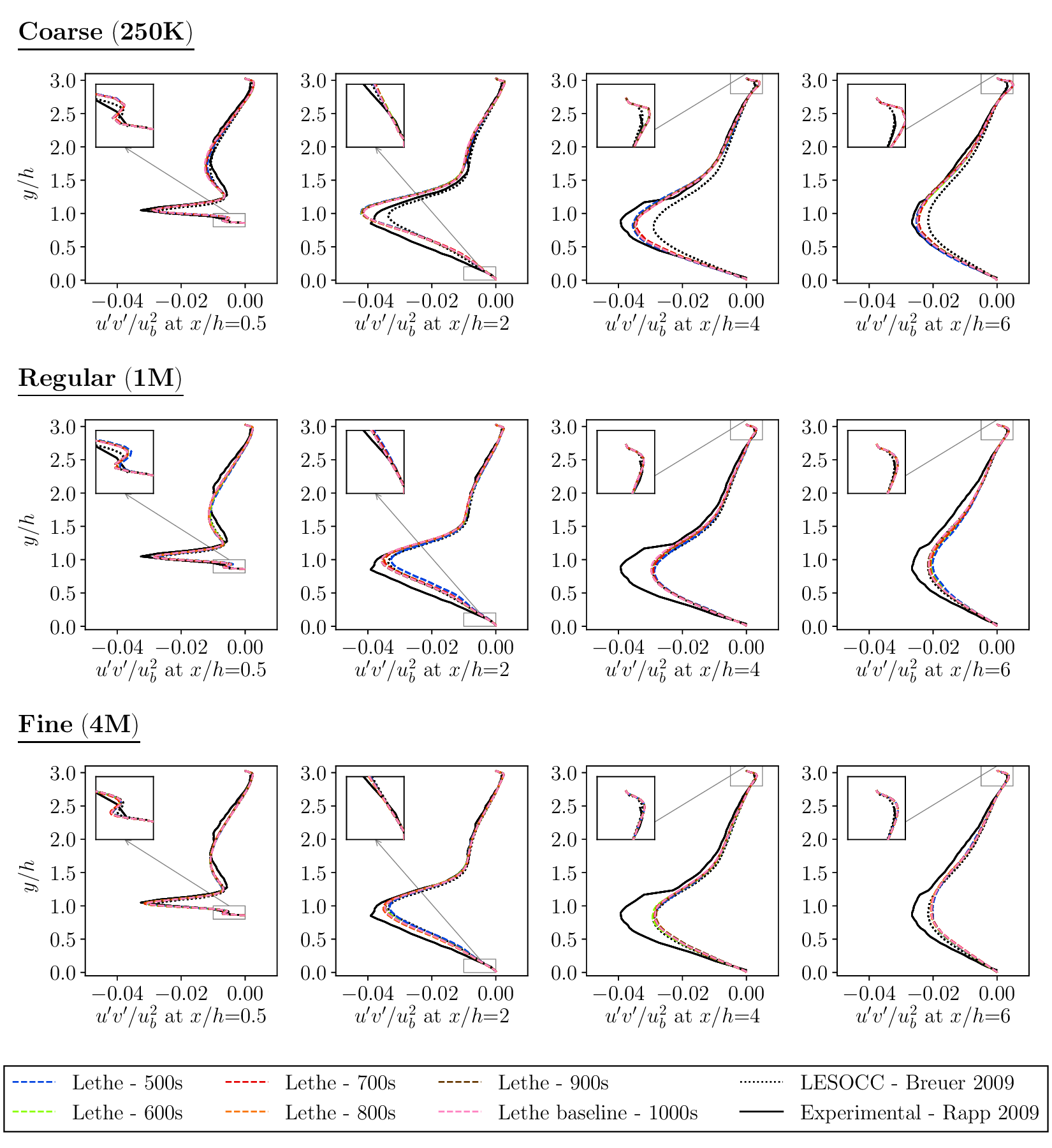}
\caption{Reynolds shear stress at different points of the geometry with $\mathrm{Re}= 5600$, compared against the benchmarks.}
\label{fig:time_averaging_reynolds_uv}
\end{figure}

According to Fig.~\ref{fig:time_averaging_velocity_0}, \ref{fig:time_averaging_reynolds_uu} and \ref{fig:time_averaging_reynolds_uv}, the results for all the meshes are average independent. The difference between the results for different time-averaging period is minimal in all cases. The largest differences can be observed near to the walls due to the zoom-in plots and it is slightly more evident when taking a look to the Reynolds stresses. To answer the question of how the time-averaging affects the prediction of the reattachment point, this value was extracted at varying averaging times for each mesh and plotted, as per the method by \cite{Krank2018}. This method plots the reattachment points against the averaging time (in number of flows through times). The error $e$ is evaluated using the following expression $e = \pm c/\sqrt{T_f}$, where $c$ is a manually defined constant and $T_f$ is the number of flows through times. Since the reattachment point oscillates due to the turbulence, a final reattachment point is set to a value that the reattachment points extracted tend towards. The constant $c$ is then manually adjusted so that the error is as small as possible with all the extracted reattachment points lying within the error range. The error bandwidth hence decreases as the averaging time increases. The reattachment points converge very quickly, with the expected oscillation at different averaging times being within a very small error range. 

\begin{figure}[ht]
\centering
\includegraphics[scale=0.5]{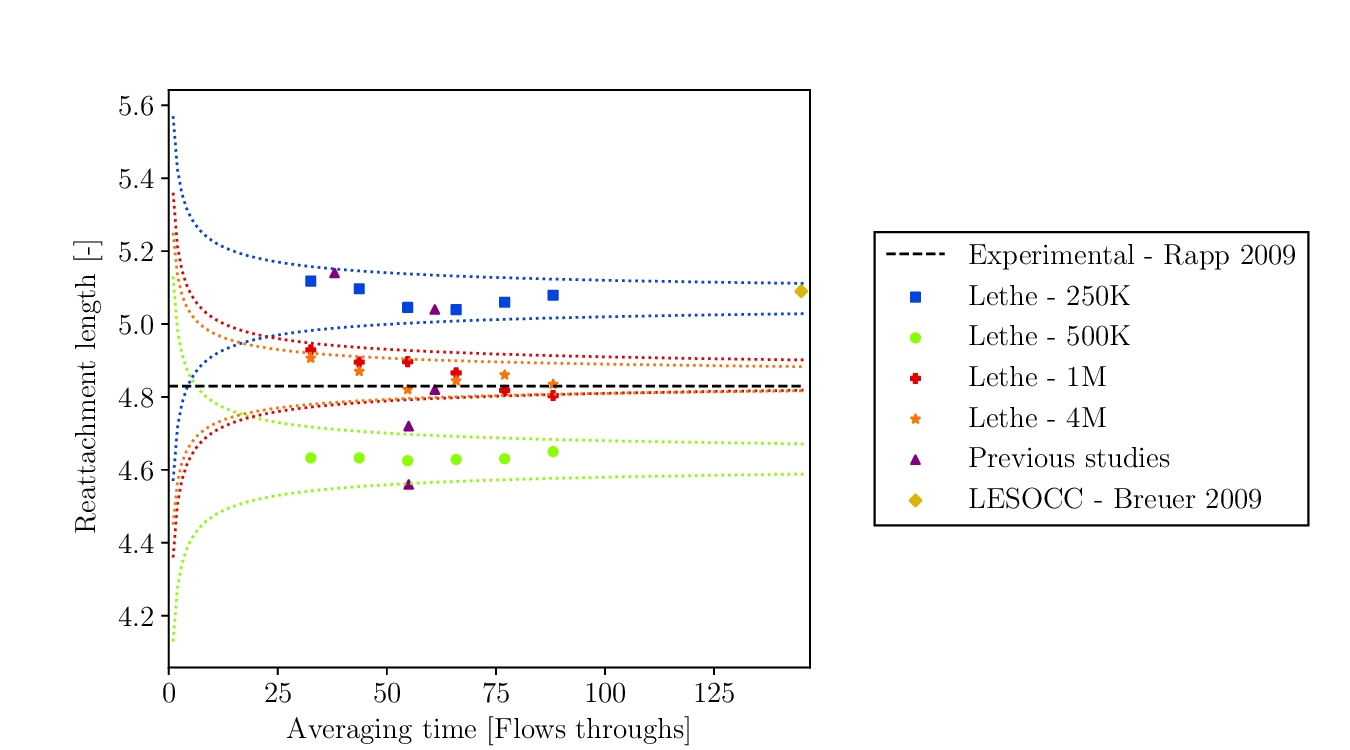}
\caption{Reattachment point extracted at different averaging times for the coarse (250K), intermediate (500K), fine (1M) and very fine (4M) meshes, where the dotted line shows the error in the reattachment point over time.}
\label{fig:averaging_reattachment_plot}
\end{figure}

The results in Fig.~\ref{fig:averaging_reattachment_plot} show that as we refine the mesh, the reattachment points approach the experimental reattachment point value by \cite{Rapp2009}. For this plot, we also added the results for an intermediate mesh with 500K cells, since it allows us to see that the results are not monotonically approaching the experimental value as the mesh is refined, which is in agreement with the oscillatory nature of the physical phenomena. The reattachment point obtained by \cite{Breuer2009} is in the error bandwidth of the reattachment points obtained using the coarse mesh in this study. In the literature, \cite{Krank2018} used this methodology to compare two configurations: a DNS simulation with 65K cells and $33.6$M DoFs and an URDNS simulation with 65K cells and $22.5$M DoFs. For the latter, they observed that the reattachment point was closer to the experimental value by \cite{Rapp2009} but away from the DNS simulation. This study shows a very clear trend towards the experimental reattachment point as we increase the mesh resolution along with a reduction of the error bandwidth. It can also be observed, that again, the bandwidth of all the meshes is reduced significantly after the averaging for $\SI{500}{\second}$ and $\SI{600}{\second}$, which reassures that a minimum of $\SI{700}{\second}$ is required to accurately predict specific quantities.

\subsection{Higher Reynolds numbers} \label{subsec:highreynolds}

As mentioned in Section \ref{sec:previous_studies}, results with higher Reynolds numbers are available in literature, therefore, we tested our ILES method with SUPG/PSPG stabilization to see how well it can predict important flow properties at $\mathrm{Re} = 10600$ and $\mathrm{Re} = 37000$. For this, three meshes are considered: very coarse, coarse and intermediate. A time step of $\Delta t = \SI{0.025}{\second}$ and a time average of $\SI{800}{\second}$ are used in accordance with the findings of the previous sections. 

\begin{figure}[ht]
\centering
\includegraphics[scale=0.7]{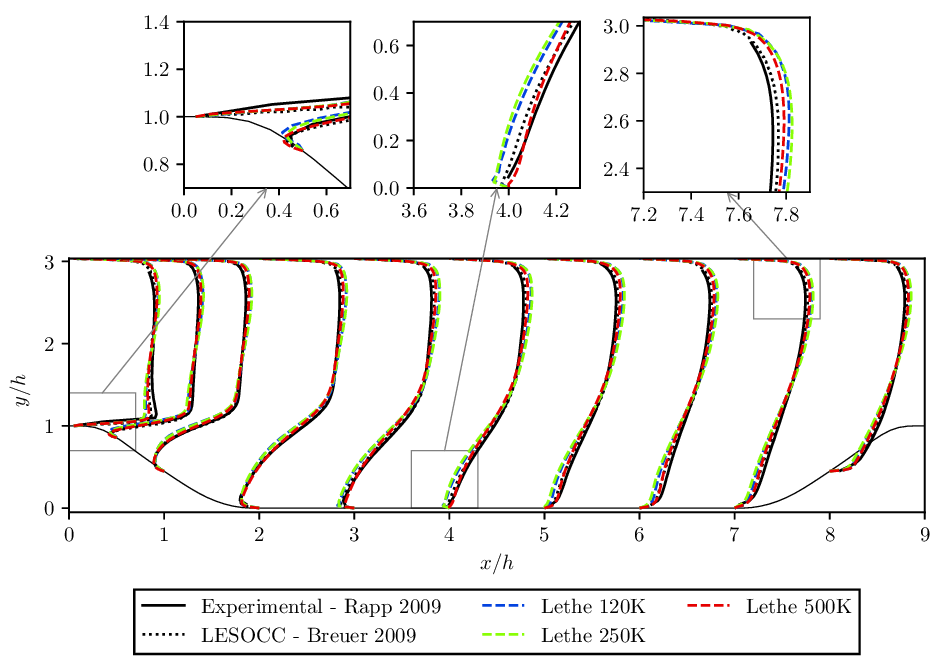}
\caption{Average velocity in the $x$ direction throughout the geometry at $\mathrm{Re} = 10600$, compared against the benchmarks. The profiles are scaled by a factor of 0.8 for ease of visualisation.}
\label{fig:high_order_velocity_10600}
\end{figure}

\begin{figure}[ht]
\centering
\includegraphics[scale=0.7]{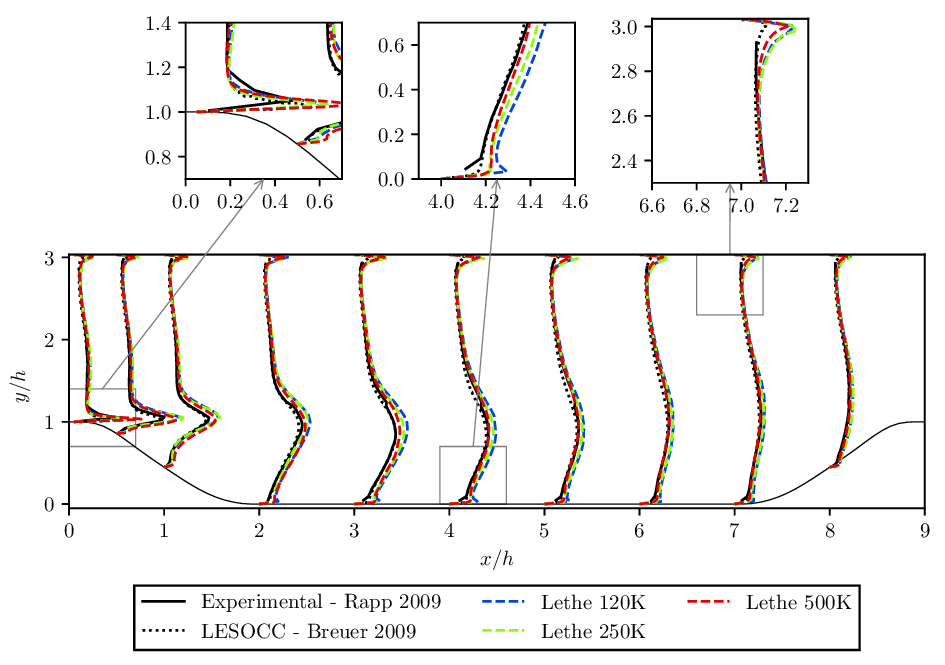}
\caption{Reynolds normal stress in the $x$ direction throughout the geometry at $\mathrm{Re} = 10600$, compared against the benchmarks. The profiles are scaled by a factor of 5 for ease of visualisation.}
\label{fig:high_order_reynolds_10600}
\end{figure}

The results for $\mathrm{Re} = 10600$ are compared to experimental and computational data by \cite{Rapp2009} and \cite{Breuer2009}, respectively. In the case of $\mathrm{Re} = 10600$, a good accuracy of the average velocity in the $x$-direction is obtained for all meshes, with the intermediate mesh having closer results to the experimental data in the bulk of the flow and near to the wall (see Fig.~\ref{fig:high_order_velocity_10600}). The reattachment points obtained for this case using the different meshes, starting from the very coarse to the intermediate are $x/h = 4.54, 4.90$ and $4.01$. According to the results obtained by Rapp ($x/h = 4.21$), the point of reattachment moves upstream with increasing Reynolds number as the recirculation zone flattens, which is what we observe in our results. In addition, the first two values are very near to the reattachment point obtained by Breuer et al. ($x/h = 4.69$). The prediction of the reattachment point is more sensitive to the time averaging period, hence it is possible that a larger period is required to obtain more accurate results in the case of this Reynolds number. For the Reynolds stresses (see Fig.~\ref{fig:high_order_reynolds_10600}) all the meshes tend to overpredict the stresses throughout the channel, again with the intermediate mesh obtaining closer results to the experimental data. 

\begin{figure}[ht]
\centering
\includegraphics[scale=0.7]{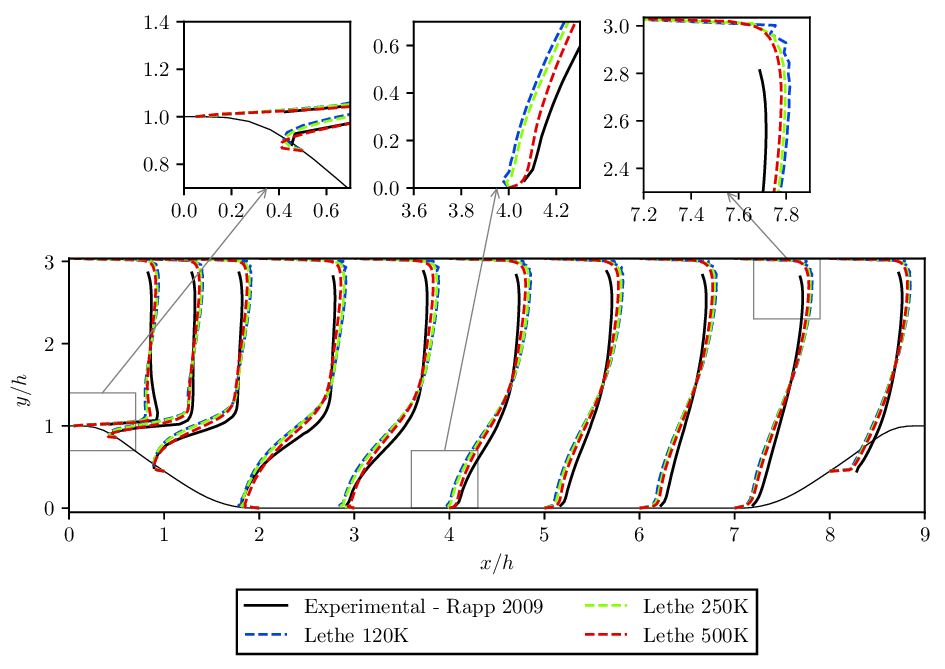}
\caption{Average velocity in the $x$ direction throughout the geometry at $\mathrm{Re} = 37000$, compared against the benchmarks. The profiles are scaled by a factor of 0.8 for ease of visualisation.}
\label{fig:high_order_velocity_37000}
\end{figure}

\begin{figure}[ht]
\centering
\includegraphics[scale=0.7]{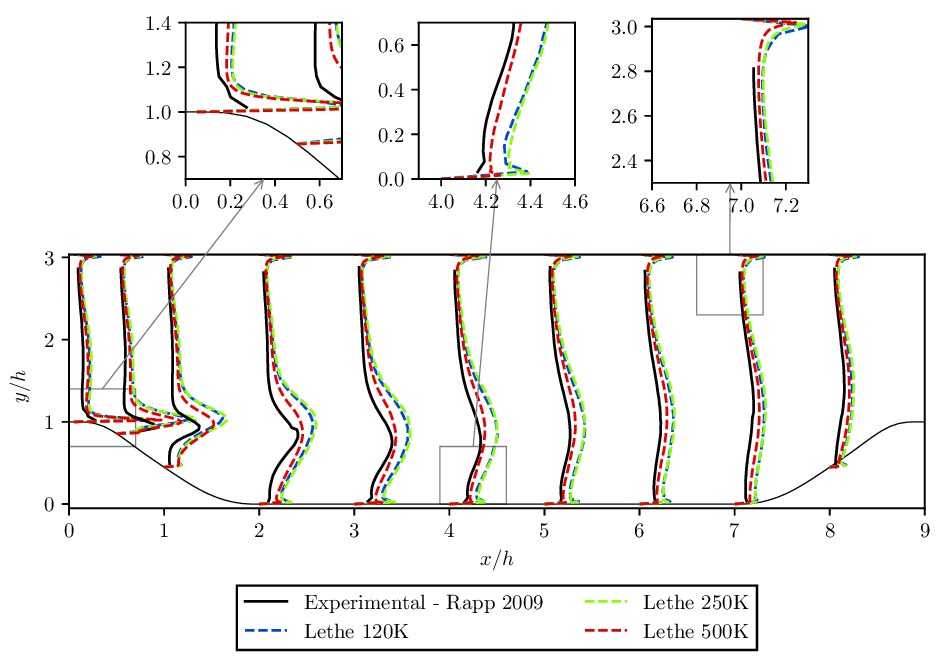}
\caption{Reynolds normal stress in the $x$ direction throughout the geometry at $\mathrm{Re}= 37000$, compared against the benchmarks. The profiles are scaled by a factor of 5 for ease of visualisation.}
\label{fig:high_order_reynolds_37000}
\end{figure}

The results for the case with $\mathrm{Re} = 37000$ were only compared to the experimental benchmark as Breuer et al. did not conduct any study using LESOCC for this Reynolds number. We see a similar trend on the estimation of the average velocity and the Reynolds stresses, however, with a higher discrepancy of the overall results in the bulk of the flow and near to the wall. In Fig.~\ref{fig:high_order_velocity_37000}, there is an underprediction of the velocity in the lower region of the channel and an overprediction on the upper region, having a higher difference near to the walls. A higher difference between the results of the different meshes can be observed in the results for the Reynolds stresses (see Fig.~\ref{fig:high_order_reynolds_37000}), where they are all overpredicted, however, the intermediate mesh is the closest one to experimental results. As pointed out by \cite{Rapp2009}, to accurately predict the high near-wall peak obtained with this Reynolds number is of utter importance to have a correct prediction of Reynolds stresses at the windward side of the hill.

The reattachment points obtained in this case were $x/h = 4.20, 4.09$ and $3.49$ for the three meshes from very coarse to intermediate. Experimentally, the reattachment point was determined to be equal to $3.76$ \cite{Rapp2009}. Therefore, refining the mesh, seems to help to have a better resolution of the flow properties allowing the reattachment point to move upstream, while introducing less dissipation. However, as in the previous case, further refining the mesh might lead to less time-dependent results and better accuracy of the predictions. This section demonstrates the capabilities of this ILES approach to simulate higher Reynolds numbers for turbulent flows with complex characteristics and the need of understanding the influence of all the parameters involved in the FEM formulation to be able to improve the predictions and understand phenomena that are not easy to observe experimentally. 

\section{Conclusion} \label{sec:conclusion}
 Multiple simulations for flow over periodic hills at $\mathrm{Re} = 5600$ have been completed using a stabilised finite element ILES approach implemented in the open-source software Lethe. The use of this numerical method for modelling this simulation case was validated by comparison to two previous studies, meeting the first aim of this study. In general, the quality of the prediction depends on three factors: the mesh, the time step (along with the type of time-stepping scheme) and the time averaging period used to obtain the turbulent statistics. We demonstrated that it is possible to simulate these kinds of flows using coarse meshes for the prediction of average quantities or specific values, such as the reattachment points, but special attention needs to be taken when choosing the time step as it significantly affects the predictions when using a stabilised approach. When a finer mesh is used, along with an implicit time-stepping scheme, larger time steps (higher CFL values) than those typically used in periodic hills studies with explicit schemes can be used without losing accuracy both in the bulk of the flow and in the region near to the wall. 
 
The results for the reattachment points extracted from different simulations using different meshes approach the value given by the experimental benchmark as the mesh is refined. The values for a $1$M cells mesh and a $4$M mesh are very close, indicating that mesh-independent results were obtained. The method was also tested for $\mathrm{Re} = 10600$ and $\mathrm{Re} = 37000$ using very coarse meshes. While the ILES approach used in this study obtains accurate results for the velocity for $\mathrm{Re} = 10600$, the Reynolds stresses are highly affected near to the separation and post-reattachment zones. The results have higher discrepancies at $\mathrm{Re} = 37000$, where significant differences are observed not only at the wall but also in the bulk of the flow. These results could be further improved by further reducing the size of the cells.

To conclude, this study has not only provided a greater understanding of how the numerical parameters affect the simulation results for the periodic hills case, but also shows that ILES methods are able to provide very good solutions for these types of complex turbulent flows with coarse and fine meshes that are coarser than the ones used in literature. It also highlights the advantages of an implicit time-stepping scheme over an explicit one in the context of stabilised methods. The ILES methods are promising for practical simulations as they provide accuracy, do not require the calibration of a subgrid scale model and can reduce the computational effort, in terms of mesh size and degrees of freedom of the numerical system, in comparison to traditional LES approaches.

\backmatter



\bmhead{Acknowledgments}
The authors would like to acknowledge the financial support from the Natural Sciences and Engineering Research Council of Canada (NSERC) through the RGPIN-2020-04510 Discovery grant. The authors would also like to acknowledge technical support and computing time provided by the Digital Research Alliance of Canada and Calcul Québec.





\begin{appendices}
\section{Geometry of hills} \label{app:polynomials}
The 6 polynomials used to describe the shape of the hill are:
\begin{enumerate}
    \item For $x \in [0; 0.3214h]$: \\
    $y(x) = \mathrm{min}(h; h + 0h x + 2.420h \times 10^{-4} x^2 - 7.588h \times 10^{-5} x^3)$
    \item For $x \in [0.3214h; 0.5h]$:\\
    $y(x) = 0.8955h + 3.484h \times 10^{-2} x - 3.629h \times 10^{-3} x^2 + 6.749h \times 10^{-5} x^3$
    \item For $x \in [0.5h; 0.7143h]$:\\
    $y(x) = 0.9213h + 2.931h \times 10^{-2} x - 3.234h \times 10^{-3} x^2 + 5.809h \times 10^{-5} x^3$
    \item For $x \in [0.7143h; 1.071h]$:\\
    $y(x) = 1.445h - 4.927h \times 10^{-2} x + 6.950h \times 10^{-4} x^2 - 7.394h \times 10^{-6} x^3$
    \item For $x \in [1.071h; 1.429h]$:\\
    $y(x) = 0.6401h + 3.123h \times 10^{-2} x - 1.988h \times 10^{-3} x^2 + 2.242h \times 10^{-5} x^3$
    \item For $x \in [1.429h; 1.929h]$:\\
    $y(x) = \mathrm{max}(0; 2.0139h - 7.180h \times 10^{-2} x + 5.875h \times 10^{-4} x^2 + 9.553h \times 10^{-7} x^3)$
\end{enumerate}

\section{Viscous sublayer estimation} \label{app:y_plus}

\begin{figure}[h!]
\centering
\includegraphics[scale=0.6]{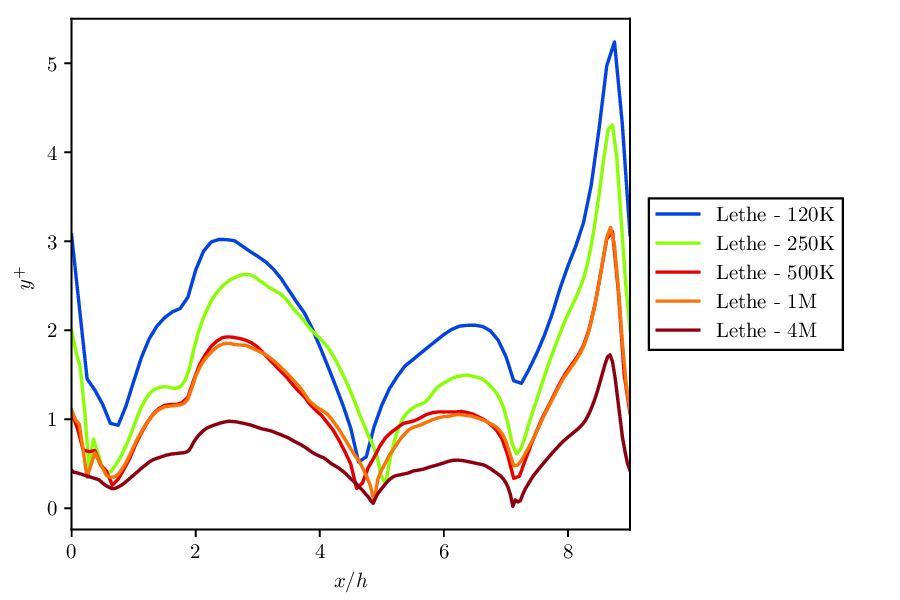}
\caption{Distribution of $y^{+}$ along the lower wall for the different meshes used at $\mathrm{Re}=5600$. This was calculated for an average time of 800 s and a time step of 0.025.}
\label{fig:yplus_5600}
\end{figure}

\begin{table}[h!]
\begin{center}
\begin{minipage}{0.9\textwidth}
\caption{Maximum and average $y^{+}$ for the different meshes at $\mathrm{Re}=5600$. This was calculated for an average time of 800 s and a time step of 0.025.}
\begin{tabular*}{\textwidth}{@{\extracolsep{\fill}}lcc@{\extracolsep{\fill}}}
\toprule%
& \multicolumn{2}{@{}c@{}}{\textbf{$y^{+}$}}  \\\cmidrule{2-3}%
\textbf{Mesh} & \textbf{Average} & \textbf{Maximum} \\
\midrule
120K              & 2.15          & 5.23  \\
250K              & 1.66        & 4.30  \\
500K              & 1.17          & 3.10  \\
1M             & 1.16          & 3.15  \\
4M              & 0.61          & 1.73  \\
\botrule
\end{tabular*}
\end{minipage}
\end{center}
\end{table}

\begin{figure}[h!]
\centering
\includegraphics[scale=0.6]{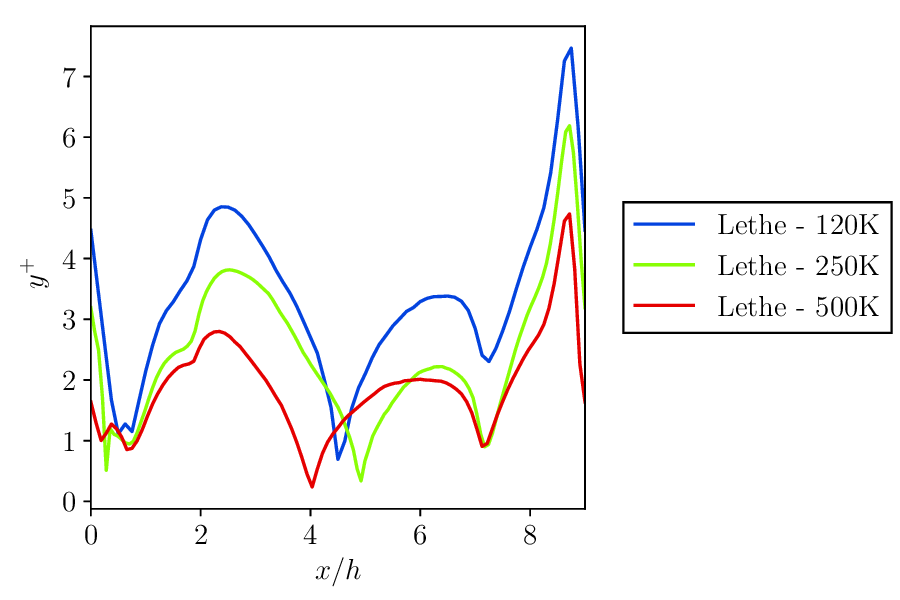}
\caption{Distribution of $y^{+}$ along the lower wall for the different meshes used at $\mathrm{Re}=10600$. This was calculated for an average time of 800 s and a time step of 0.025.}
\label{fig:yplus_10600}
\end{figure}

\begin{table}[h!]
\begin{center}
\begin{minipage}{0.9\textwidth}
\caption{Maximum and average $y^{+}$ for the different meshes at $\mathrm{Re}=10600$. This was calculated for an average time of 800 s and a time step of 0.025.}
\begin{tabular*}{\textwidth}{@{\extracolsep{\fill}}lcc@{\extracolsep{\fill}}}
\toprule%
& \multicolumn{2}{@{}c@{}}{\textbf{$y^{+}$}}  \\\cmidrule{2-3}%
\textbf{Mesh} & \textbf{Average} & \textbf{Maximum} \\
\midrule
120K              & 3.34          & 7.46  \\
250K              & 2.44        & 6.18  \\
500K              & 1.89          & 4.73  \\
\botrule
\end{tabular*}
\end{minipage}
\end{center}
\end{table}

\begin{figure}[h!]
\centering
\includegraphics[scale=0.6]{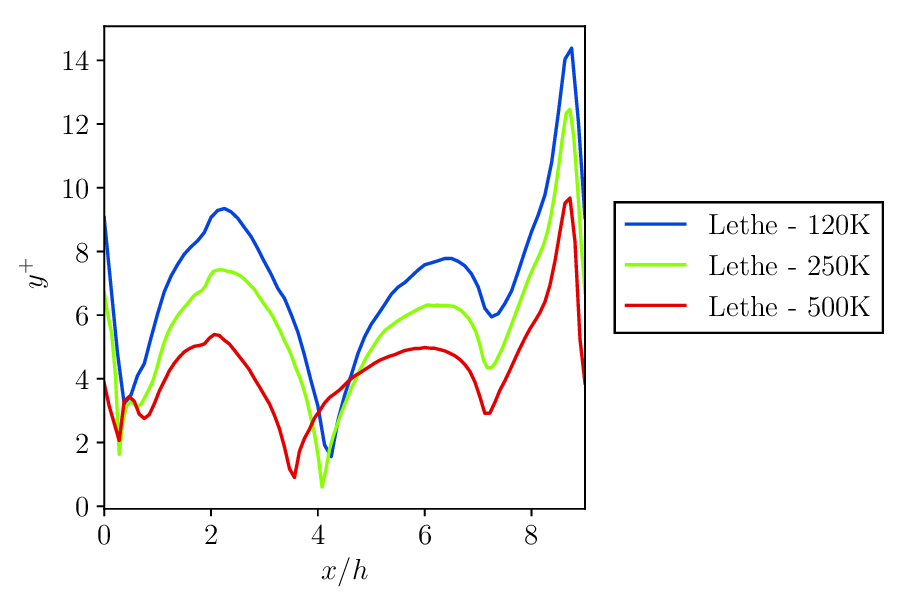}
\caption{Distribution of $y^{+}$ along the lower wall for the different meshes used at $\mathrm{Re}=37000$. This was calculated for an average time of 800 s and a time step of 0.025.}
\label{fig:yplus_37000}
\end{figure}

\begin{table}[h!]
\begin{center}
\begin{minipage}{0.9\textwidth}
\caption{Maximum and average $y^{+}$ for the different meshes at $\mathrm{Re}=37000$. This was calculated for an average time of 800 s and a time step of 0.025.}
\begin{tabular*}{\textwidth}{@{\extracolsep{\fill}}lcc@{\extracolsep{\fill}}}
\toprule%
& \multicolumn{2}{@{}c@{}}{\textbf{$y^{+}$}}  \\\cmidrule{2-3}%
\textbf{Mesh} & \textbf{Average} & \textbf{Maximum} \\
\midrule
120K              & 7.06          & 14.38  \\
250K              & 5.73        & 12.46  \\
500K              & 4.35          & 9.67  \\
\botrule
\end{tabular*}
\end{minipage}
\end{center}
\end{table}

\end{appendices}


\bibliography{main}

\begin{thebibliography}{}
\providecommand{\doi}[1]{\url{https://doi.org/#1}}
\bibcommenthead

\bibitem [\protect \citeauthoryear {%
Arndt%
\ \protect \BOthers {.}}{%
Arndt%
\ \protect \BOthers {.}}{%
{\protect \APACyear {2021}}%
}]{%
dealII93}
\APACinsertmetastar {%
dealII93}%
\begin{APACrefauthors}%
Arndt, D.%
, Bangerth, W.%
, Blais, B.%
, Fehling, M.%
, Gassm{\"o}ller, R.%
, Heister, T.%
\BDBL {}Zhang, J.%
\end{APACrefauthors}%
\unskip\
\newblock
\APACrefYearMonthDay{2021}{}{}.
\newblock
{\BBOQ}\APACrefatitle {The \texttt{deal.II} Library, Version 9.3} {The
  \texttt{deal.II} library, version 9.3}.{\BBCQ}
\newblock
\APACjournalVolNumPages{J. Numer. Math.}{29}{3}{171--186}.
\newblock
\begin{APACrefURL} {https://doi.org/10.1515/jnma-2021-0081} \end{APACrefURL}
\newblock

\newblock

\PrintBackRefs{\CurrentBib}

\bibitem [\protect \citeauthoryear {%
Arndt%
\ \protect \BOthers {.}}{%
Arndt%
\ \protect \BOthers {.}}{%
{\protect \APACyear {2022}}%
}]{%
dealII94}
\APACinsertmetastar {%
dealII94}%
\begin{APACrefauthors}%
Arndt, D.%
, Bangerth, W.%
, Feder, M.%
, Fehling, M.%
, Gassm{\"o}ller, R.%
, Heister, T.%
\BDBL {}Wells, D.%
\end{APACrefauthors}%
\unskip\
\newblock
\APACrefYearMonthDay{2022}{}{}.
\newblock
{\BBOQ}\APACrefatitle {The \texttt{deal.II} Library, Version 9.4} {The
  \texttt{deal.II} library, version 9.4}.{\BBCQ}
\newblock
\APACjournalVolNumPages{J. Numer. Math.}{30}{3}{231-246}.
\newblock
\begin{APACrefURL} {https://doi.org/10.1515/jnma-2022-0054} \end{APACrefURL}
\newblock

\newblock

\PrintBackRefs{\CurrentBib}

\bibitem [\protect \citeauthoryear {%
Balakumar%
\ \BBA {} Park%
}{%
Balakumar%
\ \BBA {} Park%
}{%
{\protect \APACyear {2015}}%
}]{%
Balakumar2015}
\APACinsertmetastar {%
Balakumar2015}%
\begin{APACrefauthors}%
Balakumar, P.%
\BCBT {}\ \BBA {} Park, G.I.%
\end{APACrefauthors}%
\unskip\
\newblock
\APACrefYearMonthDay{2015}{}{}.
\newblock
{\BBOQ}\APACrefatitle {DNS/LES Simulations of Separated Flows at High Reynolds
  Numbers} {Dns/les simulations of separated flows at high reynolds
  numbers}.{\BBCQ}
\newblock
 \APACrefbtitle {49th AIAA Fluid Dynamics Conference.} {49th aiaa fluid
  dynamics conference.}
\newblock
\begin{APACrefURL} {https://doi.org/10.2514/6.2015-2783} \end{APACrefURL}
\PrintBackRefs{\CurrentBib}

\bibitem [\protect \citeauthoryear {%
Beck%
\ \protect \BOthers {.}}{%
Beck%
\ \protect \BOthers {.}}{%
{\protect \APACyear {2014}}%
}]{%
Beck2014}
\APACinsertmetastar {%
Beck2014}%
\begin{APACrefauthors}%
Beck, A.D.%
, Bolemann, T.%
, Flad, D.%
, Frank, H.%
, Gassner, G.J.%
, Hindenlang, F.%
\BCBL {} Munz, C\BHBI D.%
\end{APACrefauthors}%
\unskip\
\newblock
\APACrefYearMonthDay{2014}{}{}.
\newblock
{\BBOQ}\APACrefatitle {High-order discontinuous Galerkin spectral element
  methods for transitional and turbulent flow simulations} {High-order
  discontinuous galerkin spectral element methods for transitional and
  turbulent flow simulations}.{\BBCQ}
\newblock
\APACjournalVolNumPages{Int. J. Numer. Methods Fluids}{76}{8}{522-548}.
\newblock
\begin{APACrefURL} {https://doi.org/10.1002/fld.3943} \end{APACrefURL}
\newblock

\newblock

\PrintBackRefs{\CurrentBib}

\bibitem [\protect \citeauthoryear {%
Benocci%
\ \BBA {} Pinelli%
}{%
Benocci%
\ \BBA {} Pinelli%
}{%
{\protect \APACyear {1990}}%
}]{%
Benocci1990}
\APACinsertmetastar {%
Benocci1990}%
\begin{APACrefauthors}%
Benocci, C.%
\BCBT {}\ \BBA {} Pinelli, A.%
\end{APACrefauthors}%
\unskip\
\newblock
\APACrefYearMonthDay{1990}{}{}.
\newblock
{\BBOQ}\APACrefatitle {The role of the forcing term in the large eddy
  simulation of equilibrium channel flow} {The role of the forcing term in the
  large eddy simulation of equilibrium channel flow}.{\BBCQ}
\newblock
 \APACrefbtitle {Engineering Turbulence Modeling and Experiments} {Engineering
  turbulence modeling and experiments}\ (\BPGS\ 287--296).
\newblock
\APACaddressPublisher{}{Elsevier}.
\newblock
\begin{APACrefURL} {https://eprints.ucm.es/id/eprint/21905/} \end{APACrefURL}
\newblock
\APACrefnote{International Symposium On Engineering Turbulence Modelling And
  Measurements , Dubrovnik, Yugoslavia, Sep 24-28, 1990}
\PrintBackRefs{\CurrentBib}

\bibitem [\protect \citeauthoryear {%
Blais%
\ \protect \BOthers {.}}{%
Blais%
\ \protect \BOthers {.}}{%
{\protect \APACyear {2020}}%
}]{%
blais2020lethe}
\APACinsertmetastar {%
blais2020lethe}%
\begin{APACrefauthors}%
Blais, B.%
, Barbeau, L.%
, Bibeau, V.%
, Gauvin, S.%
, El~Geitani, T.%
, Golshan, S.%
\BDBL {}Chaouki, J.%
\end{APACrefauthors}%
\unskip\
\newblock
\APACrefYearMonthDay{2020}{}{}.
\newblock
{\BBOQ}\APACrefatitle {Lethe: An open-source parallel high-order adaptative CFD
  solver for incompressible flows} {Lethe: An open-source parallel high-order
  adaptative cfd solver for incompressible flows}.{\BBCQ}
\newblock
\APACjournalVolNumPages{SoftwareX}{12}{}{100579}.
\newblock
\begin{APACrefURL} {https://doi.org/10.1016/j.softx.2020.100579}
  \end{APACrefURL}
\newblock

\newblock

\PrintBackRefs{\CurrentBib}

\bibitem [\protect \citeauthoryear {%
Blais%
\ \BBA {} Ilinca%
}{%
Blais%
\ \BBA {} Ilinca%
}{%
{\protect \APACyear {2018}}%
}]{%
Blais2018}
\APACinsertmetastar {%
Blais2018}%
\begin{APACrefauthors}%
Blais, B.%
\BCBT {}\ \BBA {} Ilinca, F.%
\end{APACrefauthors}%
\unskip\
\newblock
\APACrefYearMonthDay{2018}{}{}.
\newblock
{\BBOQ}\APACrefatitle {{Development and validation of a stabilized immersed
  boundary CFD model for freezing and melting with natural convection}}
  {{Development and validation of a stabilized immersed boundary CFD model for
  freezing and melting with natural convection}}.{\BBCQ}
\newblock
\APACjournalVolNumPages{Comput. Fluids}{}{}{}.
\newblock
\begin{APACrefURL} {https://doi.org/10.1016/j.compfluid.2018.03.037}
  \end{APACrefURL}
\newblock

\newblock

\PrintBackRefs{\CurrentBib}

\bibitem [\protect \citeauthoryear {%
Breuer%
\ \protect \BOthers {.}}{%
Breuer%
\ \protect \BOthers {.}}{%
{\protect \APACyear {2009}}%
}]{%
Breuer2009}
\APACinsertmetastar {%
Breuer2009}%
\begin{APACrefauthors}%
Breuer, M.%
, Peller, N.%
, Rapp, C.%
\BCBL {} Manhart, M.%
\end{APACrefauthors}%
\unskip\
\newblock
\APACrefYearMonthDay{2009}{}{}.
\newblock
{\BBOQ}\APACrefatitle {Flow over periodic hills – Numerical and experimental
  study in a wide range of Reynolds numbers} {Flow over periodic hills –
  numerical and experimental study in a wide range of reynolds numbers}.{\BBCQ}
\newblock
\APACjournalVolNumPages{Comput. Fluids}{38}{2}{433-457}.
\newblock
\begin{APACrefURL} {https://doi.org/10.1016/j.compfluid.2008.05.002}
  \end{APACrefURL}
\newblock

\newblock

\PrintBackRefs{\CurrentBib}

\bibitem [\protect \citeauthoryear {%
Calderer%
\ \BBA {} Masud%
}{%
Calderer%
\ \BBA {} Masud%
}{%
{\protect \APACyear {2013}}%
}]{%
Calderer2013}
\APACinsertmetastar {%
Calderer2013}%
\begin{APACrefauthors}%
Calderer, R.%
\BCBT {}\ \BBA {} Masud, A.%
\end{APACrefauthors}%
\unskip\
\newblock
\APACrefYearMonthDay{2013}{}{}.
\newblock
{\BBOQ}\APACrefatitle {{Residual-based variational multiscale turbulence models
  for unstructured tetrahedral meshes}} {{Residual-based variational multiscale
  turbulence models for unstructured tetrahedral meshes}}.{\BBCQ}
\newblock
\APACjournalVolNumPages{Comput. Methods Appl. Mech. Eng.}{254}{}{238--253}.
\newblock
\begin{APACrefURL} {http://dx.doi.org/10.1016/j.cma.2012.09.015}
  \end{APACrefURL}
\newblock

\newblock

\PrintBackRefs{\CurrentBib}

\bibitem [\protect \citeauthoryear {%
Chaouat%
\ \BBA {} Schiestel%
}{%
Chaouat%
\ \BBA {} Schiestel%
}{%
{\protect \APACyear {2013}}%
}]{%
Chaouat2013}
\APACinsertmetastar {%
Chaouat2013}%
\begin{APACrefauthors}%
Chaouat, B.%
\BCBT {}\ \BBA {} Schiestel, R.%
\end{APACrefauthors}%
\unskip\
\newblock
\APACrefYearMonthDay{2013}{}{}.
\newblock
{\BBOQ}\APACrefatitle {Hybrid RANS/LES simulations of the turbulent flow over
  periodic hills at high Reynolds number using the PITM method} {Hybrid
  rans/les simulations of the turbulent flow over periodic hills at high
  reynolds number using the pitm method}.{\BBCQ}
\newblock
\APACjournalVolNumPages{Comput. Fluids}{84}{}{279-300}.
\newblock
\begin{APACrefURL} {https://doi.org/10.1016/j.compfluid.2013.06.012}
  \end{APACrefURL}
\newblock

\newblock

\PrintBackRefs{\CurrentBib}

\bibitem [\protect \citeauthoryear {%
Z.~Chen%
}{%
Z.~Chen%
}{%
{\protect \APACyear {2010}}%
}]{%
Chen2010}
\APACinsertmetastar {%
Chen2010}%
\begin{APACrefauthors}%
Chen, Z.%
\end{APACrefauthors}%
\unskip\
\newblock
\APACrefYear{2010}.
\unskip\
\newblock
\APACrefbtitle {{Wall Modeling for Implicit Large-Eddy Simulation}} {{Wall
  Modeling for Implicit Large-Eddy Simulation}}\ \APACtypeAddressSchool
  {\BPhD}{}{Technical University of Munich}.
\unskip\
\newblock
\begin{APACrefURL} {https://mediatum.ub.tum.de/node?id=1007276}
  \end{APACrefURL}
\PrintBackRefs{\CurrentBib}

\bibitem [\protect \citeauthoryear {%
Z.L.~Chen%
\ \protect \BOthers {.}}{%
Z.L.~Chen%
\ \protect \BOthers {.}}{%
{\protect \APACyear {2014}}%
}]{%
Chen2014}
\APACinsertmetastar {%
Chen2014}%
\begin{APACrefauthors}%
Chen, Z.L.%
, Hickel, S.%
, Devesa, A.%
, Berland, J.%
\BCBL {} Adams, N.A.%
\end{APACrefauthors}%
\unskip\
\newblock
\APACrefYearMonthDay{2014}{}{}.
\newblock
{\BBOQ}\APACrefatitle {Wall modeling for implicit large-eddy simulation and
  immersed-interface methods} {Wall modeling for implicit large-eddy simulation
  and immersed-interface methods}.{\BBCQ}
\newblock
\APACjournalVolNumPages{Theor. Comput. Fluid Dyn.}{28}{}{1-21}.
\newblock
\begin{APACrefURL} {https://doi.org/10.1007/s00162-012-0286-6} \end{APACrefURL}
\newblock

\newblock

\PrintBackRefs{\CurrentBib}

\bibitem [\protect \citeauthoryear {%
{De la Llave Plata}%
\ \protect \BOthers {.}}{%
{De la Llave Plata}%
\ \protect \BOthers {.}}{%
{\protect \APACyear {2018}}%
}]{%
DelaLlavePlata2018}
\APACinsertmetastar {%
DelaLlavePlata2018}%
\begin{APACrefauthors}%
{De la Llave Plata}, M.%
, Couaillier, V.%
\BCBL {} le Pape, M.C.%
\end{APACrefauthors}%
\unskip\
\newblock
\APACrefYearMonthDay{2018}{}{}.
\newblock
{\BBOQ}\APACrefatitle {{DNS and LES of the flow over periodic hills based on a
  discontinuous Galerkin approach}} {{DNS and LES of the flow over periodic
  hills based on a discontinuous Galerkin approach}}.{\BBCQ}
\newblock
\APACjournalVolNumPages{Notes on Numerical Fluid Mechanics and
  Multidisciplinary Design}{135}{}{27--40}.
\newblock
\begin{APACrefURL} {https://doi.org/10.1007/978-3-319-60387-2\_3}
  \end{APACrefURL}
\newblock

\newblock

\PrintBackRefs{\CurrentBib}

\bibitem [\protect \citeauthoryear {%
Diosady%
\ \BBA {} Murman%
}{%
Diosady%
\ \BBA {} Murman%
}{%
{\protect \APACyear {2014}}%
}]{%
diosady2014}
\APACinsertmetastar {%
diosady2014}%
\begin{APACrefauthors}%
Diosady, L.T.%
\BCBT {}\ \BBA {} Murman, S.M.%
\end{APACrefauthors}%
\unskip\
\newblock
\APACrefYearMonthDay{2014}{}{}.
\newblock
{\BBOQ}\APACrefatitle {DNS of Flows over Periodic Hills using a
  Discontinuous-Galerkin Spectral-Element Method} {Dns of flows over periodic
  hills using a discontinuous-galerkin spectral-element method}.{\BBCQ}
\newblock
 A.I.~of Aeronautics\ \BBA {} Astronautics\ (\BEDS), \APACrefbtitle {44th AIAA
  Fluid Dynamics Conference, 16-20 June 2014, Atlanta, GA.} {44th aiaa fluid
  dynamics conference, 16-20 june 2014, atlanta, ga.}
\newblock
\APACaddressPublisher{}{American Institute of Aeronautics and Astronautics
  (AIAA)}.
\newblock
\begin{APACrefURL} {https://doi.org/10.2514/6.2014-2784} \end{APACrefURL}
\PrintBackRefs{\CurrentBib}

\bibitem [\protect \citeauthoryear {%
ERCOFTAC%
}{%
ERCOFTAC%
}{%
{\protect \APACyear {2017}}%
}]{%
ercoftac2017}
\APACinsertmetastar {%
ercoftac2017}%
\begin{APACrefauthors}%
ERCOFTAC%
\end{APACrefauthors}%
\unskip\
\newblock
\APACrefYearMonthDay{2017}{}{}.
\newblock
\APACrefbtitle {2D Periodic Hill Flow.} {2d periodic hill flow.}
\newblock
\begin{APACrefURL}
  {https://kbwiki.ercoftac.org/w/index.php?title=Abstr:2D\_Periodic\_Hill\_Flow}
  \end{APACrefURL}
\PrintBackRefs{\CurrentBib}

\bibitem [\protect \citeauthoryear {%
Fr\"ohlich%
\ \protect \BOthers {.}}{%
Fr\"ohlich%
\ \protect \BOthers {.}}{%
{\protect \APACyear {2005}}%
}]{%
Frohlich2005}
\APACinsertmetastar {%
Frohlich2005}%
\begin{APACrefauthors}%
Fr\"ohlich, J.%
, Mellen, C.P.%
, Rodi, W.%
, Temmerman, L.%
\BCBL {} Leschzinger, M.A.%
\end{APACrefauthors}%
\unskip\
\newblock
\APACrefYearMonthDay{2005}{}{}.
\newblock
{\BBOQ}\APACrefatitle {Highly resolved large-eddy simulation of separated flow
  in a channel with streamwise periodic constrictions} {Highly resolved
  large-eddy simulation of separated flow in a channel with streamwise periodic
  constrictions}.{\BBCQ}
\newblock
\APACjournalVolNumPages{J. Fluid Mech.}{526}{}{19–66}.
\newblock
\begin{APACrefURL} {https://doi.org/10.1017/S0022112004002812} \end{APACrefURL}
\newblock

\newblock

\PrintBackRefs{\CurrentBib}

\bibitem [\protect \citeauthoryear {%
Gamnitzer%
\ \protect \BOthers {.}}{%
Gamnitzer%
\ \protect \BOthers {.}}{%
{\protect \APACyear {2010}}%
}]{%
Gamnitzer2010}
\APACinsertmetastar {%
Gamnitzer2010}%
\begin{APACrefauthors}%
Gamnitzer, P.%
, Gravemeier, V.%
\BCBL {} Wall, W.A.%
\end{APACrefauthors}%
\unskip\
\newblock
\APACrefYearMonthDay{2010}{}{}.
\newblock
{\BBOQ}\APACrefatitle {{Time-dependent subgrid scales in residual-based large
  eddy simulation of turbulent channel flow}} {{Time-dependent subgrid scales
  in residual-based large eddy simulation of turbulent channel flow}}.{\BBCQ}
\newblock
\APACjournalVolNumPages{Comput. Methods Appl. Mech.
  Eng.}{199}{13-16}{819--827}.
\newblock
\begin{APACrefURL} {http://dx.doi.org/10.1016/j.cma.2009.07.009}
  \end{APACrefURL}
\newblock

\newblock

\PrintBackRefs{\CurrentBib}

\bibitem [\protect \citeauthoryear {%
Gloerfelt%
\ \BBA {} Cinnella%
}{%
Gloerfelt%
\ \BBA {} Cinnella%
}{%
{\protect \APACyear {2015}}%
}]{%
Gloerfelt2015}
\APACinsertmetastar {%
Gloerfelt2015}%
\begin{APACrefauthors}%
Gloerfelt, X.%
\BCBT {}\ \BBA {} Cinnella, P.%
\end{APACrefauthors}%
\unskip\
\newblock
\APACrefYearMonthDay{2015}{}{}.
\newblock
{\BBOQ}\APACrefatitle {Investigation of the flow dynamics in a channel
  constricted by periodic hills} {Investigation of the flow dynamics in a
  channel constricted by periodic hills}.{\BBCQ}
\newblock
 \APACrefbtitle {45th AIAA Fluid Dynamics Conference.} {45th aiaa fluid
  dynamics conference.}
\newblock
\APACaddressPublisher{}{American Institute of Aeronautics and Astronautics
  (AIAA)}.
\newblock
\begin{APACrefURL} {https://doi.org/10.2514/6.2015-2480} \end{APACrefURL}
\PrintBackRefs{\CurrentBib}

\bibitem [\protect \citeauthoryear {%
Gloerfelt%
\ \BBA {} Cinnella%
}{%
Gloerfelt%
\ \BBA {} Cinnella%
}{%
{\protect \APACyear {2019}}%
}]{%
Gloerfelt2019}
\APACinsertmetastar {%
Gloerfelt2019}%
\begin{APACrefauthors}%
Gloerfelt, X.%
\BCBT {}\ \BBA {} Cinnella, P.%
\end{APACrefauthors}%
\unskip\
\newblock
\APACrefYearMonthDay{2019}{06}{}.
\newblock
{\BBOQ}\APACrefatitle {Large Eddy Simulation Requirements for the Flow over
  Periodic Hills} {Large eddy simulation requirements for the flow over
  periodic hills}.{\BBCQ}
\newblock
\APACjournalVolNumPages{Flow Turbul. Combust.}{103}{}{55–91}.
\newblock
\begin{APACrefURL} {https://doi.org/10.1007/s10494-018-0005-5} \end{APACrefURL}
\newblock

\newblock

\PrintBackRefs{\CurrentBib}

\bibitem [\protect \citeauthoryear {%
Hay%
\ \protect \BOthers {.}}{%
Hay%
\ \protect \BOthers {.}}{%
{\protect \APACyear {2015}}%
}]{%
Hay2015}
\APACinsertmetastar {%
Hay2015}%
\begin{APACrefauthors}%
Hay, A.%
, Etienne, S.%
, Pelletier, D.%
\BCBL {} Garon, A.%
\end{APACrefauthors}%
\unskip\
\newblock
\APACrefYearMonthDay{2015}{jun}{}.
\newblock
{\BBOQ}\APACrefatitle {{Hp-Adaptive time integration based on the BDF for
  viscous flows}} {{Hp-Adaptive time integration based on the BDF for viscous
  flows}}.{\BBCQ}
\newblock
\APACjournalVolNumPages{J. Comput. Phys.}{291}{}{151--176}.
\newblock
\begin{APACrefURL} {https://doi.org/10.1016/j.jcp.2015.03.022} \end{APACrefURL}
\newblock

\newblock

\PrintBackRefs{\CurrentBib}

\bibitem [\protect \citeauthoryear {%
Hickel%
\ \protect \BOthers {.}}{%
Hickel%
\ \protect \BOthers {.}}{%
{\protect \APACyear {2008}}%
}]{%
Hickel2008}
\APACinsertmetastar {%
Hickel2008}%
\begin{APACrefauthors}%
Hickel, S.%
, Kempe, T.%
\BCBL {} Adams, N.A.%
\end{APACrefauthors}%
\unskip\
\newblock
\APACrefYearMonthDay{2008}{}{}.
\newblock
{\BBOQ}\APACrefatitle {Implicit large-eddy simulation applied to turbulent
  channel flow with periodic constrictions} {Implicit large-eddy simulation
  applied to turbulent channel flow with periodic constrictions}.{\BBCQ}
\newblock
\APACjournalVolNumPages{Theor. Comput. Fluid Dyn.}{22}{}{227–242}.
\newblock
\begin{APACrefURL} {https://doi.org/10.1007/s00162-007-0069-7} \end{APACrefURL}
\newblock

\newblock

\PrintBackRefs{\CurrentBib}

\bibitem [\protect \citeauthoryear {%
Hsu%
\ \protect \BOthers {.}}{%
Hsu%
\ \protect \BOthers {.}}{%
{\protect \APACyear {2010}}%
}]{%
Hsu2010}
\APACinsertmetastar {%
Hsu2010}%
\begin{APACrefauthors}%
Hsu, M.C.%
, Bazilevs, Y.%
, Calo, V.M.%
, Tezduyar, T.E.%
\BCBL {} Hughes, T.J.%
\end{APACrefauthors}%
\unskip\
\newblock
\APACrefYearMonthDay{2010}{}{}.
\newblock
{\BBOQ}\APACrefatitle {{Improving stability of stabilized and multiscale
  formulations in flow simulations at small time steps}} {{Improving stability
  of stabilized and multiscale formulations in flow simulations at small time
  steps}}.{\BBCQ}
\newblock
\APACjournalVolNumPages{Comput. Methods Appl. Mech.
  Eng.}{199}{13-16}{828--840}.
\newblock
\begin{APACrefURL} {http://dx.doi.org/10.1016/j.cma.2009.06.019}
  \end{APACrefURL}
\newblock

\newblock

\PrintBackRefs{\CurrentBib}

\bibitem [\protect \citeauthoryear {%
Ilinca%
\ \protect \BOthers {.}}{%
Ilinca%
\ \protect \BOthers {.}}{%
{\protect \APACyear {2019}}%
}]{%
Ilinca2019}
\APACinsertmetastar {%
Ilinca2019}%
\begin{APACrefauthors}%
Ilinca, F.%
, Yu, K.R.%
\BCBL {} Blais, B.%
\end{APACrefauthors}%
\unskip\
\newblock
\APACrefYearMonthDay{2019}{}{}.
\newblock
{\BBOQ}\APACrefatitle {{The effect of viscosity on free surface flow inside an
  angularly oscillating rectangular tank}} {{The effect of viscosity on free
  surface flow inside an angularly oscillating rectangular tank}}.{\BBCQ}
\newblock
\APACjournalVolNumPages{Comput. Fluids}{}{}{}.
\newblock
\begin{APACrefURL} {https://doi.org/10.1016/j.compfluid.2019.02.021}
  \end{APACrefURL}
\newblock

\newblock

\PrintBackRefs{\CurrentBib}

\bibitem [\protect \citeauthoryear {%
John%
}{%
John%
}{%
{\protect \APACyear {2016}}%
}]{%
John2016}
\APACinsertmetastar {%
John2016}%
\begin{APACrefauthors}%
John, V.%
\end{APACrefauthors}%
\unskip\
\newblock
\APACrefYear{2016}.
\newblock
\APACrefbtitle {Finite Element Methods for Incompressible Flow Problems}
  {Finite element methods for incompressible flow problems}.
\newblock
\APACaddressPublisher{}{Springer International Publishing}.
\newblock
\begin{APACrefURL} {https://doi.org/10.1007/978-3-319-45750-5} \end{APACrefURL}
\PrintBackRefs{\CurrentBib}

\bibitem [\protect \citeauthoryear {%
{Kornhaas}%
\ \protect \BOthers {.}}{%
{Kornhaas}%
\ \protect \BOthers {.}}{%
{\protect \APACyear {2008}}%
}]{%
Kornhaas2008}
\APACinsertmetastar {%
Kornhaas2008}%
\begin{APACrefauthors}%
{Kornhaas}, M.%
, {Sternel}, D.C.%
\BCBL {} {Sch{\"a}fer}, M.%
\end{APACrefauthors}%
\unskip\
\newblock
\APACrefYearMonthDay{2008}{}{}.
\newblock
{\BBOQ}\APACrefatitle {{Influence of Time Step Size and Convergence Criteria on
  Large Eddy Simulations with Implicit Time Discretization}} {{Influence of
  Time Step Size and Convergence Criteria on Large Eddy Simulations with
  Implicit Time Discretization}}.{\BBCQ}
\newblock
 J.~{Meyers}, B.J.~{Geurts}\BCBL {}\ \BBA {} P.~{Sagaut}\ (\BEDS),
  \APACrefbtitle {Quality and Reliability of Large-Eddy Simulations} {Quality
  and reliability of large-eddy simulations}\ (\BVOL~12, \BPG~119).
\newblock
\begin{APACrefURL} {https://doi.org/10.1007/978-1-4020-8578-9\_10}
  \end{APACrefURL}
\PrintBackRefs{\CurrentBib}

\bibitem [\protect \citeauthoryear {%
Krank%
\ \protect \BOthers {.}}{%
Krank%
\ \protect \BOthers {.}}{%
{\protect \APACyear {2018}}%
}]{%
Krank2018}
\APACinsertmetastar {%
Krank2018}%
\begin{APACrefauthors}%
Krank, B.%
, Kronbichler, M.%
\BCBL {} Wall, W.A.%
\end{APACrefauthors}%
\unskip\
\newblock
\APACrefYearMonthDay{2018}{}{}.
\newblock
{\BBOQ}\APACrefatitle {Direct Numerical Simulation of Flow over Periodic Hills
  up to ReH=10,595} {Direct numerical simulation of flow over periodic hills up
  to reh=10,595}.{\BBCQ}
\newblock
\APACjournalVolNumPages{Flow Turbul. Combust.}{101}{}{521-551}.
\newblock
\begin{APACrefURL} {https://doi.org/10.1007/s10494-018-9941-3} \end{APACrefURL}
\newblock

\newblock

\PrintBackRefs{\CurrentBib}

\bibitem [\protect \citeauthoryear {%
Li%
\ \protect \BOthers {.}}{%
Li%
\ \protect \BOthers {.}}{%
{\protect \APACyear {2015}}%
}]{%
Li2015}
\APACinsertmetastar {%
Li2015}%
\begin{APACrefauthors}%
Li, Z.%
, Zhang, Y.%
\BCBL {} Chen, H.%
\end{APACrefauthors}%
\unskip\
\newblock
\APACrefYearMonthDay{2015}{}{}.
\newblock
{\BBOQ}\APACrefatitle {A low dissipation numerical scheme for Implicit Large
  Eddy Simulation} {A low dissipation numerical scheme for implicit large eddy
  simulation}.{\BBCQ}
\newblock
\APACjournalVolNumPages{Comput. Fluids}{117}{}{233-246}.
\newblock
\begin{APACrefURL} {https://doi.org/10.1016/j.compfluid.2015.05.016}
  \end{APACrefURL}
\newblock

\newblock

\PrintBackRefs{\CurrentBib}

\bibitem [\protect \citeauthoryear {%
Lodato%
\ \BBA {} Chapelier%
}{%
Lodato%
\ \BBA {} Chapelier%
}{%
{\protect \APACyear {2019}}%
}]{%
Lodato2019}
\APACinsertmetastar {%
Lodato2019}%
\begin{APACrefauthors}%
Lodato, G.%
\BCBT {}\ \BBA {} Chapelier, J.B.%
\end{APACrefauthors}%
\unskip\
\newblock
\APACrefYearMonthDay{2019}{}{}.
\newblock
{\BBOQ}\APACrefatitle {{Evaluation of the spectral element dynamic model for
  LES on unstructured, deformed meshes}} {{Evaluation of the spectral element
  dynamic model for LES on unstructured, deformed meshes}}.{\BBCQ}
\newblock
\APACjournalVolNumPages{ERCOFTAC Series}{25}{}{39--45}.
\newblock
\begin{APACrefURL} {https://doi.org/10.1007/978-3-030-04915-7\_6}
  \end{APACrefURL}
\newblock

\newblock

\PrintBackRefs{\CurrentBib}

\bibitem [\protect \citeauthoryear {%
Mellen%
\ \protect \BOthers {.}}{%
Mellen%
\ \protect \BOthers {.}}{%
{\protect \APACyear {2000}}%
}]{%
Mellen2000}
\APACinsertmetastar {%
Mellen2000}%
\begin{APACrefauthors}%
Mellen, C.P.%
, Fröhlich, J.%
\BCBL {} Rodi, W.%
\end{APACrefauthors}%
\unskip\
\newblock
\APACrefYearMonthDay{2000}{August}{}.
\newblock
{\BBOQ}\APACrefatitle {Large-eddy simulation of the flow over periodic hills}
  {Large-eddy simulation of the flow over periodic hills}.{\BBCQ}
\newblock
 M.~Deville\ \BBA {} R.~Owens\ (\BEDS), \APACrefbtitle {16th IMACS world
  congress.} {16th imacs world congress.}
\newblock
\APACaddressPublisher{Lausanne, Switzerland}{}.
\PrintBackRefs{\CurrentBib}

\bibitem [\protect \citeauthoryear {%
Mitchell%
\ \protect \BOthers {.}}{%
Mitchell%
\ \protect \BOthers {.}}{%
{\protect \APACyear {2017}}%
}]{%
mitchell2017engauge}
\APACinsertmetastar {%
mitchell2017engauge}%
\begin{APACrefauthors}%
Mitchell, M.%
, Muftakhidinov, B.%
, Winchen, T.%
, van Schaik, B.%
, Wilms, A.%
\BCBL {}\ \BOthersPeriod {.}\end{APACrefauthors}%
\unskip\
\newblock
\APACrefYearMonthDay{2017}{}{}.
\newblock
{\BBOQ}\APACrefatitle {Engauge digitizer software} {Engauge digitizer
  software}.{\BBCQ}
\newblock
\APACjournalVolNumPages{Webpage: http://markummitchell. github.
  io/engauge-digitizer.}{11}{}{}.
\newblock

\newblock

\PrintBackRefs{\CurrentBib}

\bibitem [\protect \citeauthoryear {%
Mokhtarpoor%
\ \BBA {} Heinz%
}{%
Mokhtarpoor%
\ \BBA {} Heinz%
}{%
{\protect \APACyear {2017}}%
}]{%
Mokhtarpoor2017}
\APACinsertmetastar {%
Mokhtarpoor2017}%
\begin{APACrefauthors}%
Mokhtarpoor, R.%
\BCBT {}\ \BBA {} Heinz, S.%
\end{APACrefauthors}%
\unskip\
\newblock
\APACrefYearMonthDay{2017}{}{}.
\newblock
{\BBOQ}\APACrefatitle {Dynamic large eddy simulation: Stability via
  realizability} {Dynamic large eddy simulation: Stability via
  realizability}.{\BBCQ}
\newblock
\APACjournalVolNumPages{Phys. Fluids}{29}{10}{105104}.
\newblock
\begin{APACrefURL} {https://doi.org/10.1063/1.4986890} \end{APACrefURL}
\newblock

\newblock

\PrintBackRefs{\CurrentBib}

\bibitem [\protect \citeauthoryear {%
Mokhtarpoor%
\ \protect \BOthers {.}}{%
Mokhtarpoor%
\ \protect \BOthers {.}}{%
{\protect \APACyear {2016}}%
}]{%
Mokhtarpoor2016}
\APACinsertmetastar {%
Mokhtarpoor2016}%
\begin{APACrefauthors}%
Mokhtarpoor, R.%
, Heinz, S.%
\BCBL {} Stoellinger, M.%
\end{APACrefauthors}%
\unskip\
\newblock
\APACrefYearMonthDay{2016}{}{}.
\newblock
{\BBOQ}\APACrefatitle {{Dynamic unified RANS-LES simulations of high reynolds
  number separated flows}} {{Dynamic unified RANS-LES simulations of high
  reynolds number separated flows}}.{\BBCQ}
\newblock
\APACjournalVolNumPages{Phys. Fluids}{28}{9}{}.
\newblock
\begin{APACrefURL} {http://dx.doi.org/10.1063/1.4961254} \end{APACrefURL}
\newblock

\newblock

\PrintBackRefs{\CurrentBib}

\bibitem [\protect \citeauthoryear {%
Rapp%
}{%
Rapp%
}{%
{\protect \APACyear {2009}}%
}]{%
Rapp2009}
\APACinsertmetastar {%
Rapp2009}%
\begin{APACrefauthors}%
Rapp, C.%
\end{APACrefauthors}%
\unskip\
\newblock
\APACrefYear{2009}.
\unskip\
\newblock
\APACrefbtitle {Experimentelle Studie der turbulenten Strömung über
  periodische Hügel} {Experimentelle studie der turbulenten strömung über
  periodische hügel}\ \APACtypeAddressSchool {\BPhD}{}{Technische
  Universit\"at M\"unchen}.
\unskip\
\newblock
\begin{APACrefURL} {https://mediatum.ub.tum.de/node?id=677970} \end{APACrefURL}
\PrintBackRefs{\CurrentBib}

\bibitem [\protect \citeauthoryear {%
Rapp%
\ \BBA {} Manhart%
}{%
Rapp%
\ \BBA {} Manhart%
}{%
{\protect \APACyear {2011}}%
}]{%
Manhart2011}
\APACinsertmetastar {%
Manhart2011}%
\begin{APACrefauthors}%
Rapp, C.%
\BCBT {}\ \BBA {} Manhart, M.%
\end{APACrefauthors}%
\unskip\
\newblock
\APACrefYearMonthDay{2011}{07}{}.
\newblock
{\BBOQ}\APACrefatitle {Flow over periodic hills: An experimental study} {Flow
  over periodic hills: An experimental study}.{\BBCQ}
\newblock
\APACjournalVolNumPages{Exp. Fluids}{51}{}{247-269}.
\newblock
\begin{APACrefURL} {https://doi.org/10.1007/s00348-011-1045-y} \end{APACrefURL}
\newblock

\newblock

\PrintBackRefs{\CurrentBib}

\bibitem [\protect \citeauthoryear {%
Tezduyar%
}{%
Tezduyar%
}{%
{\protect \APACyear {1991}}%
}]{%
Tezduyar1992}
\APACinsertmetastar {%
Tezduyar1992}%
\begin{APACrefauthors}%
Tezduyar, T.E.%
\end{APACrefauthors}%
\unskip\
\newblock
\APACrefYearMonthDay{1991}{}{}.
\newblock
{\BBOQ}\APACrefatitle {Stabilized Finite Element Formulations for
  Incompressible Flow Computations} {Stabilized finite element formulations for
  incompressible flow computations}.{\BBCQ}
\newblock
 J.W.~Hutchinson\ \BBA {} T.Y.~Wu\ (\BEDS), (\BVOL~28, \BPG~1-44).
\newblock
\APACaddressPublisher{}{Elsevier}.
\newblock
\begin{APACrefURL} {https://doi.org/10.1016/S0065-2156(08)70153-4}
  \end{APACrefURL}
\PrintBackRefs{\CurrentBib}

\bibitem [\protect \citeauthoryear {%
Tezduyar%
\ \protect \BOthers {.}}{%
Tezduyar%
\ \protect \BOthers {.}}{%
{\protect \APACyear {1992}}%
}]{%
Tezduyar1992d}
\APACinsertmetastar {%
Tezduyar1992d}%
\begin{APACrefauthors}%
Tezduyar, T.E.%
, Mittal, S.%
, Ray, S.E.%
\BCBL {} Shih, R.%
\end{APACrefauthors}%
\unskip\
\newblock
\APACrefYearMonthDay{1992}{mar}{}.
\newblock
{\BBOQ}\APACrefatitle {{Incompressible flow computations with stabilized
  bilinear and linear equal-order-interpolation velocity-pressure elements}}
  {{Incompressible flow computations with stabilized bilinear and linear
  equal-order-interpolation velocity-pressure elements}}.{\BBCQ}
\newblock
\APACjournalVolNumPages{Comput. Methods Appl. Mech. Eng.}{95}{2}{221--242}.
\newblock
\begin{APACrefURL} {https://doi.org/10.1016/0045-7825(92)90141-6}
  \end{APACrefURL}
\newblock

\newblock

\PrintBackRefs{\CurrentBib}

\bibitem [\protect \citeauthoryear {%
Wang%
\ \protect \BOthers {.}}{%
Wang%
\ \protect \BOthers {.}}{%
{\protect \APACyear {2021}}%
}]{%
Wang2021}
\APACinsertmetastar {%
Wang2021}%
\begin{APACrefauthors}%
Wang, R.%
, Wu, F.%
, Xu, H.%
\BCBL {} Sherwin, S.J.%
\end{APACrefauthors}%
\unskip\
\newblock
\APACrefYearMonthDay{2021}{}{}.
\newblock
{\BBOQ}\APACrefatitle {Implicit large-eddy simulations of turbulent flow in a
  channel via spectral/hp element methods} {Implicit large-eddy simulations of
  turbulent flow in a channel via spectral/hp element methods}.{\BBCQ}
\newblock
\APACjournalVolNumPages{Phys. Fluids}{33}{3}{035130}.
\newblock
\begin{APACrefURL} {https://doi.org/10.1063/5.0040845} \end{APACrefURL}
\newblock

\newblock

\PrintBackRefs{\CurrentBib}

\bibitem [\protect \citeauthoryear {%
Xiao%
\ \protect \BOthers {.}}{%
Xiao%
\ \protect \BOthers {.}}{%
{\protect \APACyear {2020}}%
}]{%
Xiao2020}
\APACinsertmetastar {%
Xiao2020}%
\begin{APACrefauthors}%
Xiao, H.%
, Laizet, S.%
\BCBL {} Duan, L.%
\end{APACrefauthors}%
\unskip\
\newblock
\APACrefYearMonthDay{2020}{01}{}.
\newblock
{\BBOQ}\APACrefatitle {Flows Over Periodic Hills of Parameterized Geometries: A
  Dataset for Data-Driven Turbulence Modeling From Direct Simulations} {Flows
  over periodic hills of parameterized geometries: A dataset for data-driven
  turbulence modeling from direct simulations}.{\BBCQ}
\newblock
\APACjournalVolNumPages{Comput. Fluids}{}{}{104431}.
\newblock
\begin{APACrefURL} {https://doi.org/10.1016/j.compfluid.2020.104431}
  \end{APACrefURL}
\newblock

\newblock

\PrintBackRefs{\CurrentBib}

\end{thebibliography}


\end{document}